\documentclass[twocolumn,aps,pre,floatfix,superscriptaddress,showpacs]{revtex4}
\usepackage{graphicx}
\usepackage{bm}
\usepackage{epsfig}

\begin{document}
\title{Static and dynamic lengthscales 
in a simple glassy plaquette model}

\author{Robert L. Jack} 
\affiliation{Rudolf Peierls Centre for
Theoretical Physics, University of Oxford, 1 Keble Road, Oxford, OX1
3NP, UK}

\author{Ludovic Berthier}
\affiliation{Laboratoire des Collo\"{\i}des, Verres et Nanomat\'eriaux, 
Universit\'e Montpellier II and UMR 5587 CNRS, 34095 Montpellier 
Cedex 5, France}

\author{Juan P. Garrahan}
\affiliation{School of Physics and Astronomy, University of
Nottingham, Nottingham, NG7 2RD, UK}

\begin{abstract}
We study static and dynamic spatial correlations in a two-dimensional
spin model with four-body plaquette interactions and standard Glauber
dynamics by means of analytic arguments and Monte Carlo simulations.
We study in detail the dynamical behaviour which becomes glassy at low
temperatures, due to the emergence of effective kinetic constraints in
a dual representation where spins are mapped to plaquette variables.
We study the interplay between non-trivial static correlations of the
spins and the dynamic `four-point' correlations usually studied in the
context of supercooled liquids.  We show that slow dynamics is
spatially heterogeneous due to the presence of diverging lengthscales
and scaling, as is also found in kinetically constrained models.  This
analogy is illustrated by a comparative study of a froth model where
the kinetic constraints are imposed.
\end{abstract}


\maketitle

\section{Introduction}

Recently there has been considerable interest in the extent to which
the slow dynamics of glass-forming liquids may be understood as a
result of constraints on the kinetic properties of a system, rather
than on its thermodynamics~\cite{GCTheory,Berthier2003,steve4}.  It
has been shown that kinetically constrained models with trivial
thermodynamic properties~\cite{FAModel,EastModel,review} show a
slowing down at low temperature, accompanied by the stretched
exponential relaxation and dynamically heterogeneous behaviour
characteristic of glass-formers.  The diverging timescales arise from
dynamical fixed points at zero temperature~\cite{Whitelam2004,steve},
associated with diverging dynamical lengthscales.

However the only degrees of freedom in these models are
phenomenological `mobility fields'~\cite{canaries}.  Excitations in
the mobility field represent regions of the glass-former where motion
is possible.  Presumably the energy barriers that prevent relaxation
in the system are smaller than average in these regions, but there
remain questions as to how these degrees of freedom are related to
other physical properties of the glass-forming systems. In particular,
is local mobility associated with some (possibly complicated) local
static ordering? Is it always necessary to consider dynamic
correlators?  Is a mapping from interacting particles to a mobility
field feasible?

An interesting class of models in which mobility fields emerge
naturally from more familiar interacting degrees of freedom was
investigated in Refs.~\cite{Lipowski1997,Garrahan2002}.  In this
particular case, mobile regions are associated with defects in the
spin fields and these defects have a density that vanishes at zero
temperature, accompanied by a divergence in the relaxation time of the
system. These models therefore provide a specific, yet informative
example of effective kinetic constraints arising from simple dynamical
rules in conjunction with familiar Hamiltonians with multi-spin
interactions with no quenched disorder.

In this paper we investigate correlations in the two-dimensional
plaquette model~\cite{Lipowski1997,Garrahan2002,Buhot2002,Espriu2004}.
The thermodynamic properties of the model are those of non-interacting
point-like excitations.  However, when these energetic properties are
combined with a simple spin-flip dynamics, the dynamical evolution at
low temperatures is controlled by large energy barriers.  The
relaxation times diverge in an Arrhenius manner consistent with the
behaviour of strong glasses~\cite{Angell1995}. This is accompanied by
a diverging lengthscale associated with dynamic correlations. Our
study therefore illustrates how a mobility field
might arise from a physical interacting spin system. The resulting
dynamic
behaviour is very similar to the physics of kinetically constrained
models as we show by comparatively studying a two-dimensional kinetic
model inspired by the physics of covalent froths~\cite{Davison}. The
major difference is that in the latter case dynamics is directly
expressed in terms of the fundamental excitations and kinetic
constraints are therefore imposed by hand.  We show that the dynamical
correlations in both models can be explained in terms of freely
diffusing excitations.

In section~\ref{sec:model} we define the models, study static
correlations and the representation in terms of diffusing excitations.
In Section~\ref{sec:dynamic} we study dynamic correlations and Section
\ref{sec:conc} contains our conclusions.

\section{Plaquette and froth models}
\label{sec:model}

\subsection{The plaquette model}

The Hamiltonian of the plaquette model
reads~\cite{Lipowski1997,Garrahan2002}
\begin{equation}
H =  - \frac{1}{2} \sum_{i,j=1}^{L-1} \sigma_{ij} \sigma_{i+1,j} \sigma_{i,j+1}
\sigma_{i+1,j+1},
\label{equ:H_spins}
\end{equation}
where the $\{\sigma_{ij}\}$ are Ising spins on a two-dimensional
square lattice with free boundary conditions.  We specify simple
spin-flip dynamics with rates given by Glauber probabilities.  We
denote the linear size of the lattice by $L$, so that there are
$N=L^2$ spins in the system.  To demonstrate that the thermodynamic
properties of this model are trivial, we make the change to plaquette
variables, $\{p_{ij}\}$, defined as
\begin{equation}
p_{ij} = \sigma_{ij} \sigma_{i+1,j} \sigma_{i,j+1} \sigma_{i+1,j+1}.
\label{equ:pij}
\end{equation}
The plaquette variables form a dual representation of the spin system.
For a system with $N$ spins, the partition sum over these $N$
variables is equivalent to summing over $(L-1)^2$ independent
plaquettes, with a final summation over $(2L-1)$ spins that lie on two
orthogonal boundaries of the system.

The result is that excited plaquettes, that is those with $p_{ij}=-1$,
are uncorrelated in space because the Hamiltonian becomes trivial in
the plaquette representation,
\begin{equation}
H = - \frac{1}{2} \sum_{i,j=1}^{L-1} p_{ij}.
\label{ham2}
\end{equation}  
Excitations have a density $(e^{\beta}+1)^{-1}$ where $\beta$ is the
inverse temperature. Also, the free summation over boundary spins
means that all states have a degeneracy of $2^{2L-1}$. This arises
from the symmetry of the Hamiltonian under flipping all of the spins
in any row or column of the square lattice. A result of the symmetry
is that only correlation functions that are invariant under this
symmetry can take finite values, since there is no spontaneous
symmetry breaking in this model. For example, the two-point correlator
\begin{equation}
\langle \sigma_{ij} \sigma_{i+x,j+y}\rangle  = \delta_{x,0} \delta_{y,0}
\label{equ:2pt_correl}
\end{equation}
It is clear that this correlator must vanish if either $x$ or $y$
is non-zero, since
flipping either row $i+x$ or column $j+y$ changes
the sign of the correlation function 
without changing the energy of the state.

Interestingly the plaquette model also describes~\cite{Jack8v2004cm}
the paramagnetic phase of the eight vertex
model~\cite{BaxterBook}. The additional presence of ordered states in
the eight vertex model makes it a suitable analogy for a glass-former
with a thermodynamic melting transition.  It turns out that the
plaquette model remains a good description for the `supercooled'
states below $T_c$, see Ref.~\cite{Jack8v2004cm}.

\subsection{Static lengthscales}
\label{sec:statics}

\begin{figure}
\epsfig{file=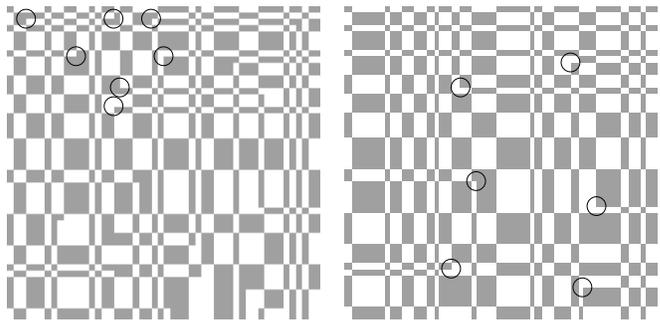,width=\columnwidth}
\caption{Typical spin configurations at $\beta=5$ (left) and $\beta=7$
(right). At the lower temperature we identify the excited plaquettes
with circles, which appear as `corners' in the spin field. At the
higher temperature we only highlight some of the defects because they
are more numerous.  The smallest visible lengthscale is the lattice
spacing. All two-point correlations vanish, even between adjacent
sites.}
\label{fig:para}
\end{figure}

We now discuss the aspects of the static properties of the plaquette
model that are relevant to the physics of supercooled liquids and
kinetically constrained models.  We mentioned above that the
concentration of excited plaquettes in the model is
$(e^\beta+1)^{-1}$. The interesting behaviour in the model occurs at
$T<1$ (so $\beta>1$), in which these excitations are dilute.  We
therefore define
\begin{equation}
c=e^{-\beta}\ll 1.
\end{equation}
Two typical configurations of the spins at two different low
temperatures are shown in Fig.~\ref{fig:para}. It is clear that even
though all two-point functions vanish by the symmetry discussed above,
the system is not in a typical paramagnetic state.  The axes of the
underlying lattice are apparent, and their influence is felt even at
relatively large lengthscales.

To understand static correlations we note that excited plaquettes are
topological defects, in the sense that while their energy cost is
localised, removing a single excited plaquette requires changes in the
spin field over large distances.  This is apparent
from the behaviour of the four-point correlation function
\begin{equation}
C_4^{\rm stat}(x,y) = \langle \sigma_{ij} \sigma_{i+x,j} \sigma_{i,j+y}
\sigma_{i+x,j+y} \rangle,
\label{c4stat}
\end{equation}
which is independent of $i$ and $j$ in a translationally invariant
system (or sufficiently far away from the free boundaries of
a finite one).
We have added the superscript `stat' to emphasize that we
consider here static correlations, as opposed to the `four-point'
dynamic correlators studied in Section~\ref{sec:dynamic}.  The
symmetries of the Hamiltonian mean that (\ref{c4stat}) is the only
non-trivial four-point function that can have a finite
value. Symmetries also ensure the vanishing of all the disconnected
parts of that function.  

Now, the value of the combination
$\sigma_{ij} \sigma_{i+x,j} \sigma_{i,j+y} \sigma_{i+x,j+y}$ is given
by the parity of the number of excited plaquettes in the rectangular
region defined by the four spins. To see this, we may take $x,y>0$
without loss of generality, and write
\begin{equation}
\sigma_{ij} \sigma_{i+x,j} \sigma_{i,j+y} \sigma_{i+x,j+y}
=\prod_{x'=0}^{x-1} \prod_{y'=0}^{y-1} p_{i+x',j+y'}
\end{equation}
where the $p_{ij}$ were defined in (\ref{equ:pij}). This rewriting 
relies only on the fact that the spins are Ising variables
so that $\sigma_{ij}^2=1$. The product is over all
plaquettes in the rectangular region and is $+1$ if there is an
even number of excited plaquettes and $-1$ if there is an odd number.
Thus, if there is only one excited
plaquette in the rectangle, then the product of the four spins is equal
to $-1$,
regardless of $x$ and $y$. This is a signature of a topological
defect.

When excited plaquettes are dilute the number of
excitations in a rectangular region of area $|xy|$
is a Poissonian random
variable with mean $c|xy|$. The probability that
this variable takes an even value is 
$(1+e^{-2c|xy|})/2$.
The four-point static correlator is therefore 
\begin{equation}
C_4^{\rm stat}(x,y) = e^{-2c|xy|}, 
\label{equ:c4stat_result}
\end{equation}
from which one can identify the static correlation lengthscale
$\xi_4^{\rm stat} =c^{-1/2}=e^{\beta/2}$. Physically this lengthscale
measures the typical spacing between excited plaquettes. Clearly this
lengthscale diverges at $T=0$ where the concentration of defects
vanishes.

It is interesting to note that in a standard kinetically constrained
model where the Hamiltonian is directly expressed in terms of the
relevant excitations, the distance between defects is trivially related
to their concentration. In the present case, the study of simple
structural correlators, e.g. two-point structure factors, does not
provide information on these relevant lengthscales; they are only
revealed when higher-order correlators are considered. More physically
this means that the typical size of black and white domains in
Fig.~\ref{fig:para} does not depend on temperature while the
concentration of circles does, so that by naively looking at the
interacting spins one would get the wrong impression that there is no
diverging lengthscale in the system.  Taking the analogy with liquids
seriously this suggests that while the structure factor of supercooled
liquids shows no particular trend when temperature is decreased,
higher-order correlators related to some yet unknown local ordering
could reveal the presence of increasing lengthscales, as assumed in
several scenarii of the glass transition~\cite{wolynes,tarjus,mezard}.

The problem with the correlator (\ref{c4stat}) is that it is highly
specific to the model under study, which has special symmetries. It
would be more useful to define static quantities which are independent
of the model under study but still reveal the presence of growing
lengthscales.  An idea is that in the presence of diverging
lengthscales, large fluctuations can also be expected.  Let us
consider the two-point quantity,
\begin{equation}
c_2^{\rm stat} (x,y) = \frac{1}{N} 
\sum_{i,j=1}^{L} \sigma_{ij} \sigma_{i+x,j+y}.
\end{equation}
where the sum is over a finite region of an infinite 
spin system. The number of terms in the sum is $N=L^2$;
we will later take the limit of large $L$ so
as to extract a well-defined measure of the size of fluctuations.

The expectation value of $c_2^\mathrm{stat}$ is a sum of
two point correlation functions. Eq.~(\ref{equ:2pt_correl}) therefore
constrains it to vanish in the thermodynamic limit (except
in the trivial case, $x=y=0$). In other words,
$\langle c_2^{\rm stat} (x,y) \rangle = \delta_{x,0} \delta_{y,0}$.
However, even when the expectation value of $c_2^{\rm stat}$ is zero,
its fluctuations are not.  One can
therefore define the following susceptibility,
\begin{equation}
\chi_2^{\rm stat}(x,y) = N \big[ \langle c_2^{\rm stat} (x,y)^2 \rangle 
- \langle c_2^{\rm stat} (x,y) \rangle^2 \big],
\label{sum}
\end{equation}
where the disconnected part in fact vanishes by symmetry [except
at $x=y=0$ where it exactly cancels the connected part, so that
$\chi_2^\mathrm{stat}(0,0)=0$].  
The connected part is a sum of four point static expectation
values, but symmetries of the model constrain most of these to 
be zero as well. The only non-zero terms are either of
the form of (\ref{c4stat}), or else trivial (for example
$\langle \sigma_{ij}^2 \sigma_{i'j'}^2\rangle$). Assuming that
at least one of $x$ and $y$ is non-zero, we write 
out the sums and collect the non-vanishing terms, arriving at
\begin{eqnarray}
\chi_2^{\rm stat}(x,y) & = & \frac{1}{N} \sum_{i,i',j,j'=1}^L
\langle \sigma_{ij} \sigma_{i+x,j+y} \sigma_{i'j'} \sigma_{i+x,j'+y}
\rangle
\nonumber \\ & = & (1-\delta_{x,0})(1-\delta_{y,0}) 
+ \nonumber \\ & & \frac{1}{N} \sum_{i,j,m=1}^L \left[
\delta_{x,0} C_4^\mathrm{stat}(i-m,y) + \right.
\nonumber \\ & & \hspace{48pt} \left. \delta_{y,0} 
C_4^\mathrm{stat}(x,j-m) \right] 
\end{eqnarray}
We note that the first term gives a trivial value of unity for
$\chi_2^{\rm stat}(x,y)$ unless at least one of $x$ or $y$ is equal
to zero. This arises from the four point correlators with
$i=i'$ and $j=j'$; we will see shortly that 
$\chi_2^{\rm stat}(x,0)$ also approaches this trivial value in 
when $x$ becomes much longer than the relevant correlation length.

The definition of $c_2^\mathrm{stat}$ contains a dependence on the
cluster size, $L$. However, the physically relevant limit is when
$L$ is much greater than all correlation lengths in the
system; in this limit then
$\chi_2^{\rm stat}(x,y)$ converges to a finite value.
From (\ref{equ:c4stat_result}), we can identify the sums
as geometric series:
taking care not to double count the terms
with $i=i'$ or $j=j'$, we find that 
\begin{equation}
\lim_{L\to\infty} \chi_2^{\rm stat}(x,0) = \sum_{y=-\infty}^{\infty} e^{-2c|xy|}
= \coth \left( c|x| \right),
\label{equ:chi2stat}
\end{equation}
for $x\neq0$.  From these fluctuations we identify a
second static lengthscale
\begin{equation}
\xi_2^{\rm stat} = c^{-1} = e^\beta,
\label{xi2stat}
\end{equation}
which also diverges at $T=0$.  Physically this lengthscale represents
the mean distance between adjacent excited plaquettes in the same row
or column of the square lattice, and is therefore much greater than
the typical spacing between excited plaquettes, $\xi_4^{\rm stat}$.

\begin{figure}
\begin{center} \epsfig{file=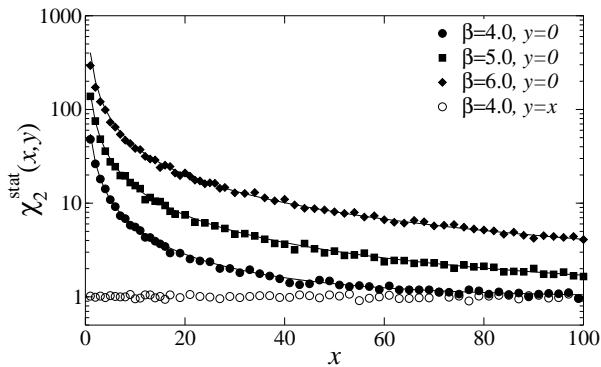, width=0.95\columnwidth}
\end{center}
\caption{Simulation data confirming the analytic results. We show
$\chi_2^\mathrm{stat}$ for $y=0$ at three different temperatures (filled
symbols);
the solid lines are the result of (\ref{equ:chi2stat}).
We also show that $\chi_2^\mathrm{stat}$ takes the trivial value of
unity when neither $x$ nor $y$ is zero. We use a cluster of $10^6$ spins
to avoid finite size effects.}
\label{fig:static_mc}
\end{figure}

In deriving Eq.~(\ref{xi2stat}) we again made use of the special
symmetries of the model, which is an undesired feature of this
treatment.  This can be
cured by noting that $\xi_2^{\rm stat}$ may be accessed through a
circular average,
\begin{equation}
\chi_2^{\rm stat}(r) \equiv 
\int_0^\infty \frac{\mathrm{d}\theta}{2\pi r} \, 
\chi^{\rm stat}(x,y) = 
\frac{2}{\pi r} \coth (cr) +1
\end{equation}
where the integral over $\theta$ is to be understood as a sum over the
integrand at values of $x,y$ such that
$r-\frac{1}{2}<\sqrt{x^2+y^2}<r+\frac{1}{2}$.  Thus the lengthscale
$\xi_2^{\rm stat}$ can indeed be extracted from fluctuations of
two-point measurements that do not depend on knowing the exact
orientation of the lattice axes or the special symmetries of the
model.

To verify these exact results, we performed static Monte Carlo
simulations to obtain $\chi_2^{\rm stat}(x,y)$ both for $y=0$ and
for the trivial case $x,y\neq0$ where the susceptibility takes the value
unity. The results are shown in figure~\ref{fig:static_mc} 

\subsection{Inherent structures and `droplets'}

We now discuss the above findings in relation to alternative approaches
to the glass transition that involve growing lengthscales of a static
nature. A problem with these approaches is that there exist no
numerical or experimental indications for the presence of such static
correlations.  However two recent papers have discussed methods to
detect growing static lengthscales in supercooled liquids that we now
discuss in the context of the plaquette model.

Bertin~\cite{bertin} suggested to consider the spatial structure of
inherent structures~\cite{IS} and studied a one-dimensional disordered
Potts model to illustrate his ideas. The procedure is as follows.
Take an equilibrium configuration of the system at finite temperature
$T$ and quench it to its inherent structure. Now fix the orientation
of the spins on the boundary of a finite portion of the system of
linear size $\ell$. Finally, minimize the energy of the finite size
system given these boundary conditions. Bertin finds that there exists
a well-defined lengthscale, $\ell^*(T)$, which is such that it is
typically possible to find a lower energy structure when $\ell >
\ell^*$, while the system is already at its ground state for smaller
sizes, $\ell < \ell^*$.

In the plaquette model, the effect of quenching to $T=0$ is that
dimers freely diffuse (see below) until they get absorbed, and then
all motion stops. The inherent structure is therefore a structure with
single defects only. On quenching from equilibrium these defects will
be randomly distributed in space, except that they are never on
adjacent sites (since that would be a dimer). So, consider a rectangular
region of the sample of linear size $\ell$. We may ask, what is the
probability that this region minimises the energy, subject to its
boundaries remaining fixed? Fixing the $4(\ell-1)$ spins that lie on
the boundaries constrains the parity of the number of defects in each
row and in each column.  Thus the (possibly degenerate) ground state
of the square region contains at most one defect in each row or
column, according to their parities. It is then obvious that when
$\ell > \ell^* \sim \xi_2^{\rm stat}$ there will typically be more
than one defect in each line and column and the inherent structure
will not coincide with the ground state of the finite size system. We
conclude therefore that Bertin's method would successfully determine
the static correlation lengthscale $\xi_2^{\rm stat}$ discussed in the
previous section.  The main difference between our model and that of
Bertin is that we cannot identify a tiling of inherent structures with
regions that minimize the energy because our defects are point-like
objects that obviously can not delimit a particular area.

In a real space description of the random first order transition it is
imagined that above its static Kauzmann transition the system is
composed of an assembly of `entropic
droplets'~\cite{wolynes}. Revisiting this idea Bouchaud and Biroli
proposed to associate a temperature dependent lengthscale, $\xi^*(T)$,
to these droplets in the following manner~\cite{jp}.  Consider an
equilibrium configuration of the system at temperature $T$.  Then
consider a finite portion of the system of linear size $\ell$ and
freeze its boundary conditions. Now let this finite size system evolve
at temperature $T$ with its boundary conditions fixed. If $\ell <
\xi^*$ the system should effectively be non-ergodic while dynamic
correlation functions would go to zero at large times for $\ell >
\xi^*$.

In the plaquette model this procedure was studied from a different
perspective in Ref.~\cite{Espriu2004} where the thermodynamics of
finite size systems was computed exactly for a particular choice of
boundary conditions.  It was found that in a finite size system there
indeed exists a `glass transition' temperature, which is manifested by
a downward jump in the specific heat which shifts to lower temperature
for larger $\ell$.  Although this calculation is not exactly the
procedure described by Bouchaud and Biroli where any sort of boundary
conditions should be studied, it is very close in spirit.  Physically,
the non-ergodic behaviour is due to the fact that when $\ell$ is too
small the system cannot rearrange without altering its boundaries.
This happens as above when lines or columns only contain one or no
defect, i.e. when $\ell < \xi_2^{\rm stat}$ and we conclude that the
procedure of Ref.~\cite{jp} would once more yield the correct
correlation lengthscale, $\xi^* \sim \xi_2^{\rm stat}$.

\subsection{Diffusing excitations in the plaquette model}

\begin{figure}
\epsfig{file=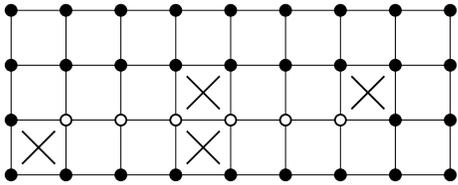,width=0.7\columnwidth}
\caption{Sketch of a dimer explaining the origin of the strongly
anisotropic dynamical correlations in the plaquette model.  Excited
plaquettes are marked with a $\times$ sign. Spins are marked by small
circles but the state of the spins is not shown. The dimer diffuses
along the $x$ direction with no energy barrier.  Open circles mark
spins that can be flipped as part of this free diffusive
process. Flipping of spins marked by closed circles costs at least
$\Delta E =2$.}
\label{fig:dimer_sketch}
\end{figure}

We now consider the dynamics of the plaquette model.  The fundamental
moves are spin-flips. When a single spin is flipped the states of the
four plaquettes surrounding that spin must all change.  Thus there are
five types of move, depending on the environment of the relevant
spin. If a spin has no adjacent excited plaquettes, then flipping that
spin incurs an energy cost $\Delta E = 4$, so these moves are
suppressed by a factor $c^4$.  Now, if a spin is adjacent to exactly
one excited plaquette, then its flips cost $\Delta E = 2$ and are
suppressed only by a factor $c^2$. The density of such sites is
approximately $c$ at low temperatures, so these flips are already more
significant than those involving $\Delta E = 4$. Finally spins
adjacent to pairs of excited plaquettes can flip without energy cost,
$\Delta E = 0$, and that those adjacent to three or four excited
plaquettes have $\Delta E<0$ and will therefore relax rapidly to the
low energy state with one or zero excited plaquettes, respectively.

The key physical point is that excited plaquettes act as sources of mobility,
since the energetic barriers to spin flips are smaller in those
regions.  This observation allows us to identify the excited
plaquettes as the excitations in the `mobility field' by analogy with
kinetically constrained models.  The square lattice of the plaquette
model has a surprising effect on the diffusion of these mobility
excitations, because pairs of adjacent excited plaquettes diffuse
freely, but are confined to one dimension, see
Fig.~\ref{fig:dimer_sketch}. As shown in 
Refs.~\cite{Buhot2002,Espriu2004}, 
this is responsible for the form of
the low temperature divergence of the relaxation time, $\tau \sim
c^{-3}$.

\subsection{The froth model}
\label{subsec:froth}

\begin{figure}
\epsfig{file=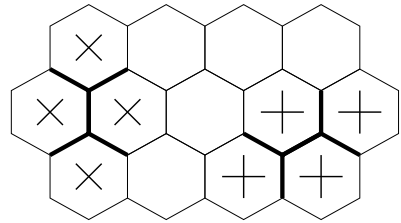,width=0.6\columnwidth}
\caption{Two possible moves in the froth model. The binary variables
$n_i$ are defined on the hexagons.  The $+$ and $\times$ signs mark
sets of four $n_i$. A single move involves flipping all four of any
such set.  The `Feynman diagram' shapes are shown in bold.}
\label{fig:feyn_diag}
\end{figure}

We now introduce a simple model that shares many features in its
dynamical behaviour with the plaquette model.  Consider a mobility
field that is defined on the plaquettes of a hexagonal lattice. Each
defect carries an energy of unity, so the Hamiltonian takes the
trivial form
\begin{equation}
H_\mathrm{froth} = -\frac{1}{2} \sum_{i=1}^N n_i,
\label{ham3}
\end{equation}
where there is a defect on site $i$ if $n_i=-1$; otherwise $n_i=+1$.
Dynamical moves involve flipping the values of four of the $n_i$ on
plaquettes that are related by a `Feynman diagram' shape, see
Fig.~\ref{fig:feyn_diag}.  This sort of model was studied in the
context of covalent froths~\cite{Davison}, so we refer to it as the
`froth model' in what follows. The similarity between Hamiltonians
(\ref{ham2}) and (\ref{ham3}) is evident but the froth model does not
have an equivalent spin representation.

Comparisons between the dynamics of the froth model and those of the
plaquette model are instructive because the only difference between
their dynamics is the structure of the underlying lattice.  However,
while the diffusing pairs of excitations in the plaquette model were
confined to one dimension as a result of the square lattice,
Fig.~\ref{fig:dimer_sketch}, the equivalent dimers of the froth model
are free to diffuse across the whole two-dimensional plane,
Fig.~\ref{fig:feyn_diag}.

\begin{figure}
\epsfig{file=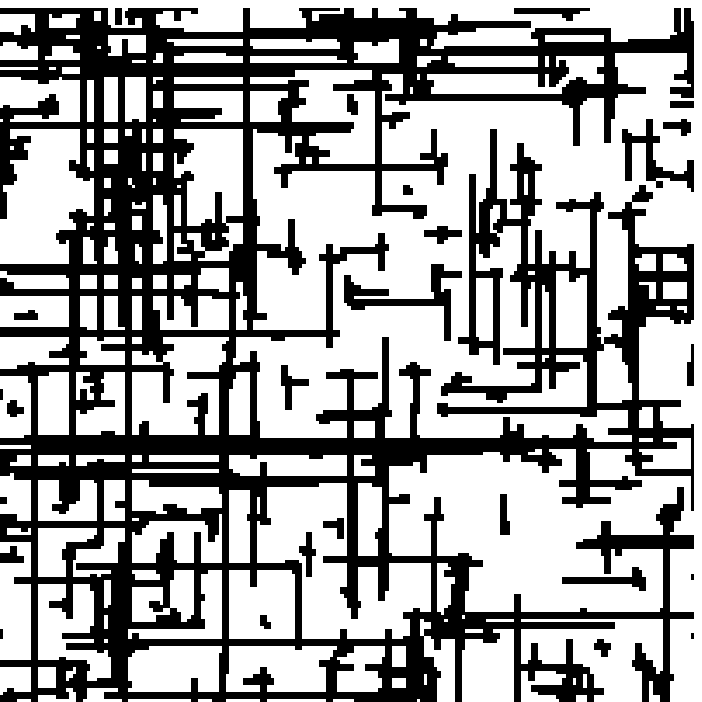,width=0.77\columnwidth}

\vspace*{.2cm}
\epsfig{file=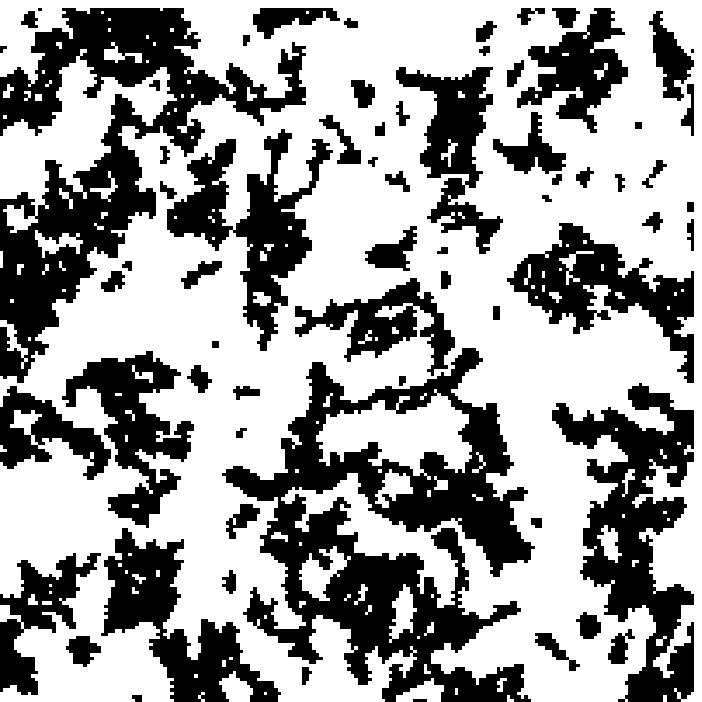,width=0.77\columnwidth}
\caption{Realisations of persistence function in the plaquette model
(top) and the froth model (bottom).  Plaquettes that have flipped
between time zero and time $t$ are coloured black. The time $t$ is
such that the fraction of black plaquettes is approximately
$0.4$. Both simulations are at $\beta=5$ and show regions of size
$200\times200$. The strong anisotropy in the top panel is purely a
result of the underlying square lattice in the plaquette model.}
\label{fig:pers_space}
\end{figure}

To first illustrate the strong effect of the choice of lattice on the
dynamics we show typical realisations of the persistence
function~\cite{Berthier2003} of each system in
Fig.~\ref{fig:pers_space}. The persistence function on site $i$ is
defined by $b_i(t,t')=1$ if the state of that site has remained
constant for all times between $t'$ and $t+t'$, otherwise
$b_i(t,t')=0$. It is clear from Fig.~\ref{fig:pers_space} that the
effect of the square lattice is to introduce strong anisotropy into
the dynamic spatial correlations.  The remainder of this paper
contains discussions of the dynamic correlations shown in
Fig.~\ref{fig:pers_space}.

\section{Dynamic correlations}
\label{sec:dynamic}

\subsection{Dynamic correlations in the plaquette model}
\label{subsec:C22}

In the previous section we introduced the persistence function to
quantify dynamics in the plaquette model.  This was a convenient way
to compare dynamics between froth and plaquette models.  However the
presence of the underlying spin field in the plaquette model means
that its dynamics is more naturally investigated in terms of the
spin-spin autocorrelation
\begin{equation}
a_{ij}(t,t') = \sigma_{ij}(t') \sigma_{ij}(t'+t).
\label{equ:ac_def}
\end{equation}
The two-time form of this operator makes it more natural than the
persistence function which depends on the spin at all times between
$t'$ and $t'+t$.

\begin{figure}
\epsfig{file=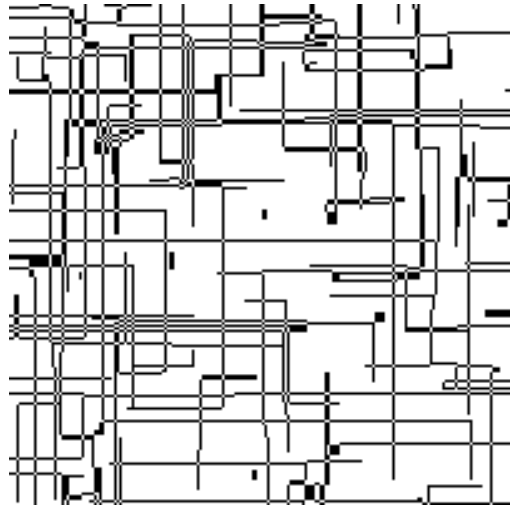,width=0.77\columnwidth}
\caption{Typical realisation of the autocorrelation in the plaquette
model. Sites with $a_{ij}=-1$ are coloured black. The inverse
temperature is $\beta=5$ and the timescale is such that $\langle
a_{ij}\rangle \simeq 0.6$.  This figure is from the same time series
as for the persistence field in Fig.~\ref{fig:pers_space} (top) but
the time $t$ is longer because most of the contributions to the
persistence are from dimers that are reabsorbed at their emission
point.}
\label{fig:ac_sketch}
\end{figure}

A typical realisation of $a_{ij}(t,t')$ is shown in
Fig.~\ref{fig:ac_sketch}. The autocorrelation function retains
slightly more information than the persistence function. If the system
makes an excursion to an excited state before returning to its initial
one, then the autocorrelation function records the fact that the
overall state has not changed, but the persistence function does
not. For this reason, the autocorrelation data shows the spatial
correlations more clearly than the persistence data.

We now investigate the anisotropic dynamic correlations in the
plaquette model more carefully. Consider the two-point, two-time
(i.e. `four-point') correlation function
\begin{eqnarray}
\tilde C_{2,2}(x,y,t) & = & \langle a_{ij}(t,t_0) a_{i+x,j+y}(t,t_0) \rangle
\nonumber\\ & &
 - \langle a_{ij}(t,t_0) \rangle \langle a_{i+x,j+y}(t,t_0) \rangle.
\end{eqnarray}
In equilibrium this function is independent of $(i,j,t_0)$ since the
system is invariant under translations in both space and time. It is
convenient to discuss also normalised correlations,
\begin{eqnarray}
C_{2,2}(x,y,t) = \frac{\tilde C_{2,2}(x,y,t)}{1 - \langle a_{ij}(t,t_0)
\rangle^2},
\end{eqnarray}
so that ${C}_{2,2}(0,0,t)=1$.

\begin{figure}
\epsfig{file=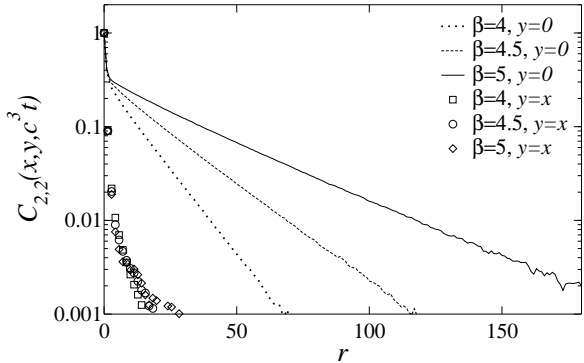,width=0.95\columnwidth}\\
\caption{Plots of $C_{2,2}(x,y,c^3 t)$ with $y=0$ and with $y=x$, as a
function of $r=\sqrt{x^2+y^2}$. Times are such that $c^3 t=0.07$ so
that $\langle a_{ij}(t,t') \rangle \simeq 0.5$.  The lengthscale for
on-axis correlations increases with decreasing temperature, but there
is no scaling for the off-axis correlator.}
\label{fig:C22}
\end{figure}

We are more particularly interested in the scaling of ${C}_{2,2}$ at
low temperatures. We find that the scaling contains contributions from
diverging length and timescales, consistent with the presence of a
dynamical critical point at $T=0$.  For example, the on-site
autocorrelation function $\langle a_{ij}(t,t')\rangle$ obeys the
scaling relation~\cite{Buhot2002} $\langle a_{ij}(t,t_0) \rangle =
f_1(c^3 t)$.  However the spatial scaling is not as simple as this
temporal relation.  It is clear from Fig.~\ref{fig:ac_sketch} that the
presence of the underlying square lattice is relevant to $C_{2,2}$
even at low temperatures.

We have investigated the function ${C}_{2,2}$ in some detail
using Monte Carlo simulations with
a continuous time algorithm~\cite{Newman-Barkema}. 
Since there is a correlation length
proportional to $1/c$ in the static properties of the system, it
is necessary to use rather large system sizes. We use periodic
boundary conditions, which require systems large enough that typical
rows and columns of the lattice contain at least two excited plaquettes.
Otherwise there are rather strong finite size effects, as observed
in~\cite{Espriu2004}. Typically we use system sizes of $10^6$ spins.
The typical error bars are of the order of the symbol sizes in 
the figures, except where noted otherwise. Time is measured in 
Monte Carlo sweeps throughout and all distances are in units
of the lattice spacing.

In Fig.~\ref{fig:C22} we show the strong correlations in ${C}_{2,2}$
along the axes of the model, and their rapid decrease away from those
axes. In order to show data at different temperatures, we use times $t$ that
scale with $c^{-3}$ so that $\langle a_{ij}(t,t')\rangle$ is the same
at each temperature.  We see that the lengthscale associated with
$C_{2,2}(x,0,t)$ grows rapidly at low temperatures while the off-axis
correlations measured by $C_{2,2}(x,x,t)$ are weak and do not depend
strongly on $\beta$.  In the representation of
Fig.~\ref{fig:ac_sketch}, ${C}_{2,2}(x,0,t)$ measures the typical
length of the rod-like objects, which gets large at small
temperatures. However, the absence of correlations away from the axes
of the lattice indicates a lack of correlation between those
objects. The microscopic origin of this behaviour is in the
one-dimensional diffusion of pairs of excited plaquettes sketched in
Fig.~\ref{fig:dimer_sketch}.

\begin{figure}
\epsfig{file=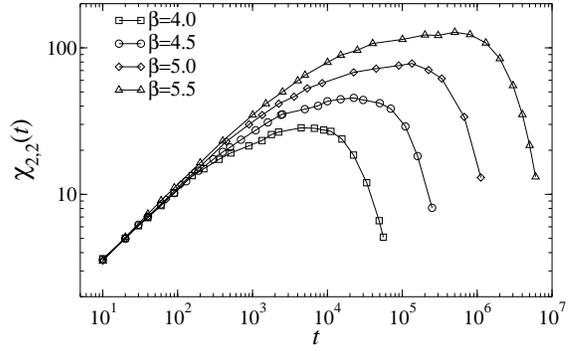,width=0.9\columnwidth}
\epsfig{file=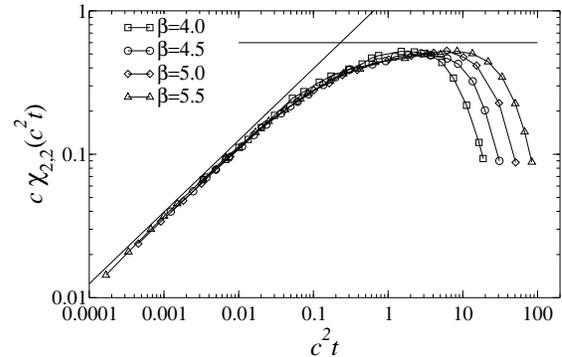,width=0.9\columnwidth}
\epsfig{file=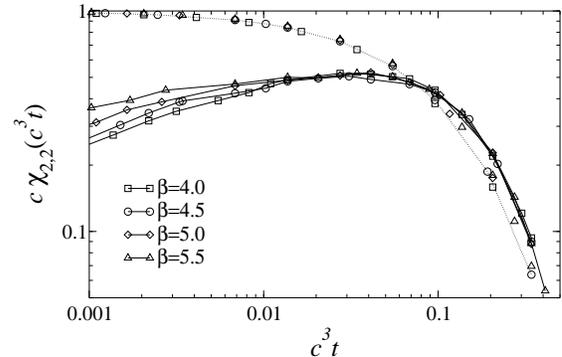,width=0.9\columnwidth}
\caption{Top: Plots of $\chi_{2,2}(t)$ at various temperatures.
Middle: Rescaling of time to $c^2 t$ showing saturation at $c^2
\tau_\mathrm{sat} \sim 1$. Lines are a power law, $t^{0.5}$,
consistent with diffusive motion, and a horizontal plateau showing
saturation at $c\chi_{2,2}\sim 0.6$.  Bottom: Rescaling of time to
$c^3 t$ showing decrease of correlations due to decreasing $\rho$
which scales in the same way as the spin-spin autocorrelation functions.
(The autocorrelation functions are
displayed as the data that collapse on the dotted line.)}
\label{fig:chi_t}
\end{figure}

We now consider the full time and space dependence of
${C}_{2,2}(x,0,t)$ in a little more detail. It is convenient to start
the discussion by focusing on the corresponding dynamical `four-point'
susceptibility~\cite{glotzer,mayer}
\begin{equation}
\chi_{2,2}(t) = \sum_{xy} {C}_{2,2}(x,y,t).
\end{equation}
This quantity is proportional both to the typical correlation area and
to the strength of the correlations. From Fig.~\ref{fig:C22} we expect
${C}_{2,2}(x,0,t) \approx \rho(t)f[x/\xi_{2,2}(t)]$ at large $x$ (see
also below) so that we may write
\begin{equation}
\chi_{2,2}(t) \approx  \rho(t) \, \xi_{2,2}(t)^{d_f}.
\label{equ:rho_xi}
\end{equation}
In this expression $d_f$ is the fractal dimension associated with the
correlations, $\xi_{2,2}(t)$ the linear size of a correlated region,
$\xi_{2,2}^{d_f}$ its area and the quantity $\rho(t)$ measures the
strength of the correlations.  We show $\chi_{2,2}(t)$ at several
temperatures in Fig.~\ref{fig:chi_t} (top).  There are three distinct
regimes in this function: a power law increase at small times; a broad
maximum whose width increases with decreasing temperature; and a rapid
decrease at larger times.

These data are easily interpreted in terms of the excitations
described in Section~\ref{sec:model}.  At short timescales the
rod-like objects shown in
in Fig.~\ref{fig:ac_sketch} grow in a diffusive manner.
These objects have $d_f=1$, and we therefore expect
$\chi_{2,2}(t) \sim (t^{1/2})^{d_f=1}$, as observed in
Fig.~\ref{fig:chi_t}. Moreover this behaviour is independent of $c$
since a pair of excited plaquettes can diffuse at no energy cost.

At larger times, dimers that have diffused can be absorbed on
encountering an isolated excited plaquette, preventing further growth
of the rods. The dynamic lengthscale therefore saturates to the mean
distance between isolated defects along one direction of the square
lattice.  This is precisely the physical content of the static
lengthscale $\xi_2^{\rm stat} \sim c^{-1}$ discussed above. Since
the process is diffusive, saturation of $\xi_{2,2}$ to
$\xi_2^{\rm stat}$ takes place at a timescale given by
\begin{equation}
\tau_\mathrm{sat} \sim c^{-2}.
\end{equation} 
Saturation at time $\tau_{\rm sat}$ is observed in 
the numerical data, as shown in the middle frame of Fig.~\ref{fig:chi_t}.

Finally, after saturation there is little change in the correlations
until the susceptibility starts to decrease for times $t\sim c^{-3}$.
Scaling in this late-time regime is governed by the time dependence of
the factor $\rho(t)$ in Eq.~(\ref{equ:rho_xi}) which starts to
decrease significantly when the rods start to overlap. When two rods
cross the site at which they intersect has flipped twice. It therefore
has $a_{ij}=+1$, unlike the rest of the rod which has
$a_{ij}=-1$. This effect takes place on the timescale set by the
autocorrelation function, which scales as $\tau \sim c^{-3} \gg
\tau_{\rm sat}$.  This behaviour is confirmed in the bottom panel of
Fig.~\ref{fig:chi_t} where the scaling of $\langle a_{ij} \rangle$ is
also shown.

\subsection{Dynamic Scaling in the plaquette model}
\label{subsec:scaling}

\begin{figure}
\epsfig{file=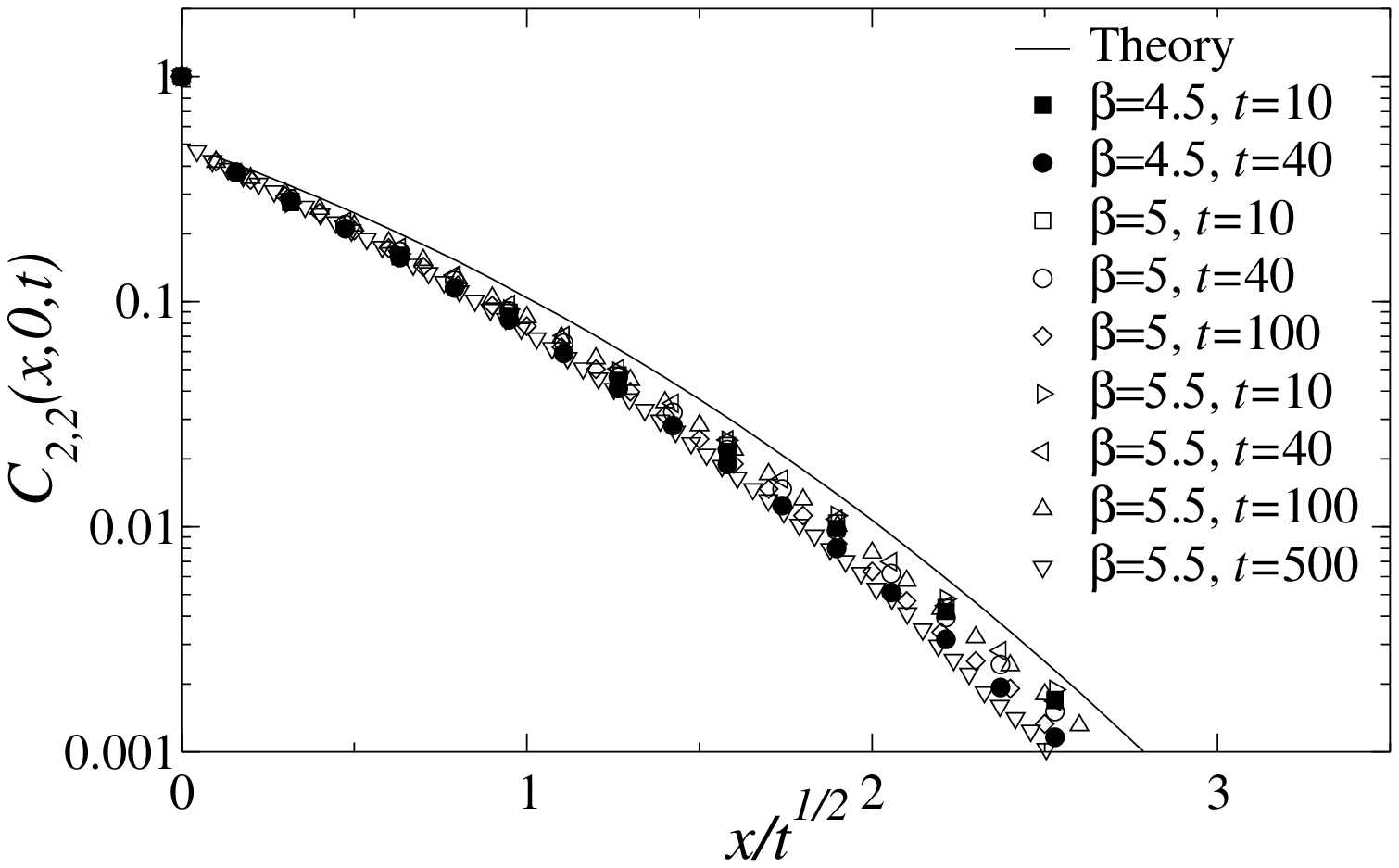,width=0.8\columnwidth}
\epsfig{file=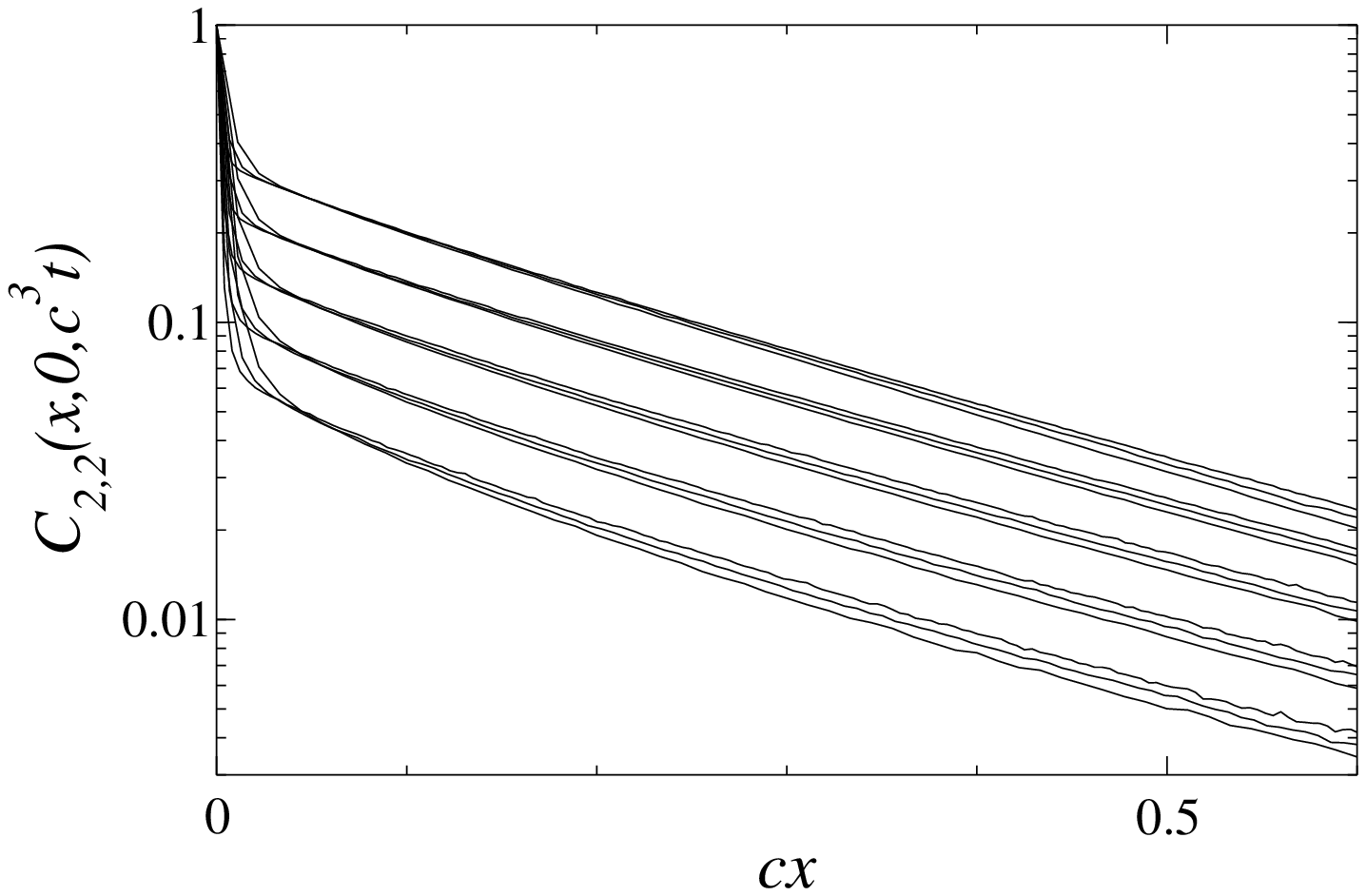,width=0.8\columnwidth}
\epsfig{file=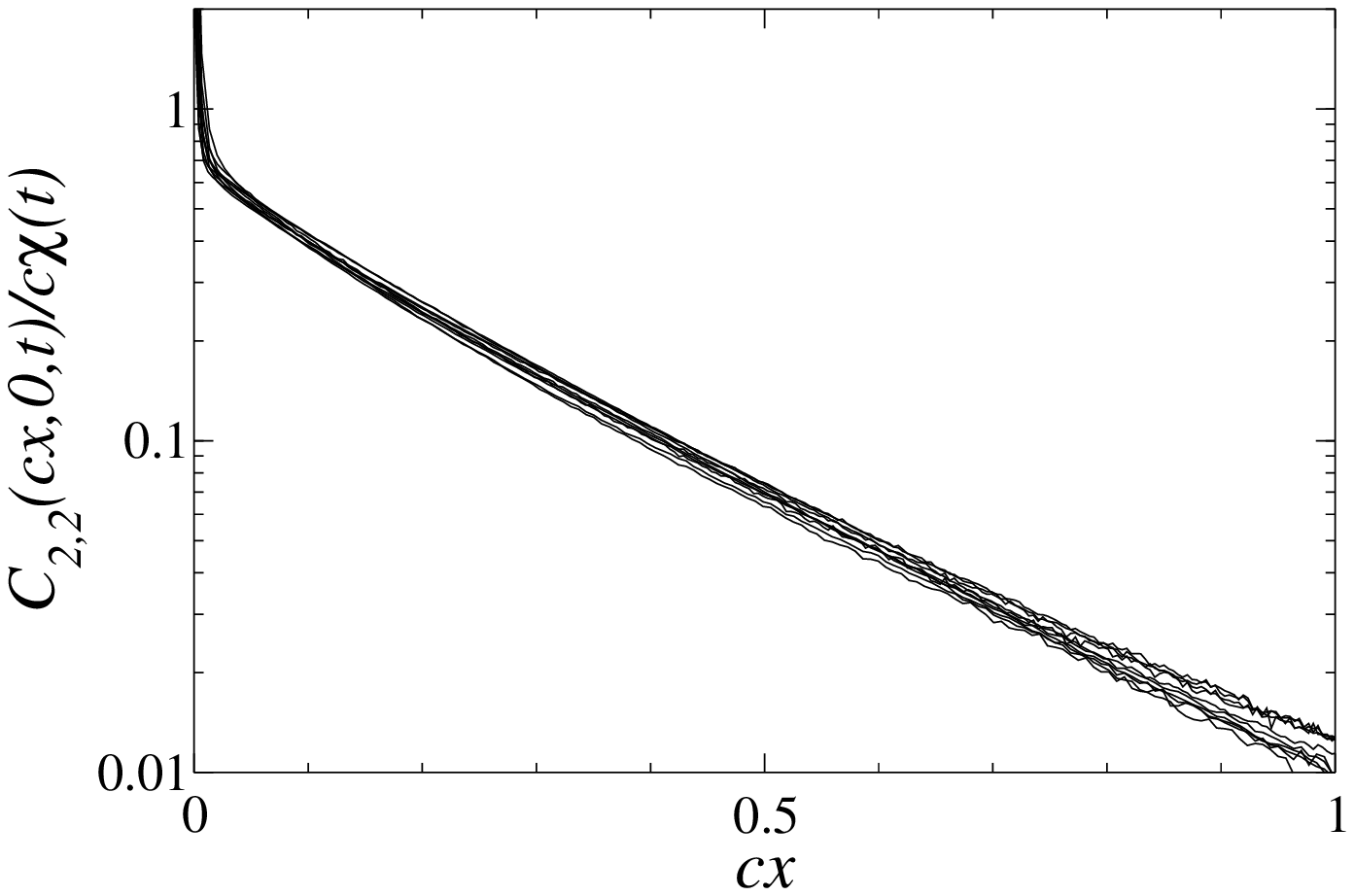,width=0.8\columnwidth}
\caption{Top: Data collapse of $C_{2,2}(x,0,t)$ on rescaling of space
by $t^{1/2}$ independently of $c$. The solid line is the prediction
for this function, assuming free (one-dimensional) diffusion of pairs
of excited plaquettes, valid for times $t \ll \tau_\mathrm{sat}$
[Eq.~(\ref{equ:c2_unsat_exact})].  Middle: Scaling of $C_{2,2}(x,0,t)$
for $t\gg \tau_\mathrm{sat}$.  We plot temperatures 
between $\beta=4$ and $\beta=5.5$ for
different scaled times $c^3 t=1,2,3,4,5$ (top to bottom). Errors
are smaller than the distances between traces. Bottom:
Collapse of the data of the middle panel using
$C_{2,2}(x,0,t)/[c\chi_{2,2}(t)]$ showing the scaling function
$f_{2c}(x)$ in Eq.~\ref{resume}.}
\label{fig:c2_dyn_scaling}
\end{figure}

We now discuss the spatial structures in $C_{2,2}(x,0,t)$, and their
scaling.  At small times, pairs of defects diffuse in one dimension.
Dynamic correlations only depend on $x/\xi_{2,2}(t) \sim x/t^{1/2}$,
and the scaling of $\chi$ is consistent with $d_f=1$.  In this time
regime we may identify a dynamical exponent, $z=2$.  At larger times,
$\tau_{\rm sat} \ll t$, the dynamical lengthscale saturates at
$\xi_{2,2}(t) \sim \xi_2^{\rm stat}$.  This saturation of the dynamic
lengthscale reflects the fact that the typical distance that a dimer
travels before being absorbed is given by $\xi_2^{\rm stat}$. We may
identify this length as a `mean free path' for dimers diffusing along
one dimensional paths.  Thus the structure of dynamical correlations
is explained in terms of the constrained one dimensional motion of
diffusing pairs of defects. The behaviour may be summarised as
\begin{equation}
\label{resume}
C_{2,2}(x,0,t) = \left\{ \begin{array}{lc} f_{2a}(x^2/t), & c^2 t \ll 1, \\
f_{2b}(c^3t) f_{2c}(cx), & c^2 t \gg 1. \end{array} \right.
\end{equation}

We illustrate this behaviour in Fig.~\ref{fig:c2_dyn_scaling}.  The
function $f_{2a}$ is shown in the top panel, along with our
theoretical prediction for it [Eq.~(\ref{equ:c2_unsat_exact}),
discussed below].  In the middle panel, we show the spatial
correlations at various times that are all greater than the saturation
time.  The spatial dependence of the correlations is not changing,
while their strength gets weaker and scales with the relaxation time
$c^{-3}$.  Finally, we may normalise the correlations by their spatial
integral as a way to extract the scaling function $f_{2c}(cx)$ in the
bottom panel of Fig.~\ref{fig:c2_dyn_scaling}.

Note that these scaling plots contain no free parameters. There
remain small deviations from scaling arising from the presence of
the lattice at small distances and from the fact 
that the two timescales $\tau_\mathrm{sat}$
and $\tau$ are not infinitely separated. However
Fig.~\ref{fig:c2_dyn_scaling} is clear evidence that the dynamical
correlations are well-described by independent random walkers in one
dimension, with a mean lifetime of approximately $\tau_\mathrm{sat}$.

We may evaluate exactly the function $f_{2a}(x^2/t)$, which gives the
value of $C_{2,2}(x,0,t)$ for times much smaller than the saturation
time. The four spin function $a_{ij}(t,t_0) a_{i+x,j}(t,t_0)$ takes
the value $-1$ if either (1) a dimer starts between the sites $(i,j)$
and $(i+x,j)$ and diffuses out of that region, or (2) a dimer starts
outside that region and diffuses into it, or (3) a dimer diffuses
along the $y$ direction, flipping one of the spins, or (4) some more
complicated process occurs, involving two dimers or the absorption
or emission of a dimer by an isolated excited plaquette.

Assuming that the dimers diffuse freely then the
first three processes mentioned above are easy to analyse by
manipulation of the diffusion equation in one dimension.
We arrive at the following result for $C_{2,2}(x,0,t)$:
\begin{equation}
f_{2a}(x^2/t) = \frac{1}{2}\left[ e^{-x^2/2t} - \sqrt{\pi} \frac{x}{\sqrt{2t}}
\,\hbox{erfc}\left(\frac{x}{\sqrt{2t}}\right)\right] + \mathcal{O}(c^2)
\label{equ:c2_unsat_exact}
\end{equation}
where $\mathrm{erfc}(x)\equiv (2/\sqrt{\pi})\int_x^\infty
\mathrm{d}y\, e^{-y^2}$.

In deriving this result we assumed that only dimers present at $t=t_0$
affect the spins, and that these dimers diffuse freely throughout 
the system.  This assumption breaks down when
the dimers present in the initial state start to encounter isolated
defects, where they are absorbed.  This is the physics behind the
collapse in the middle panel of figure~\ref{fig:chi_t}, and happens
when $c^2 t$ reaches a value of order unity.
The absorption of the dimers cuts off the growth of 
the dynamical lengthscale, leading to shorter range correlations than
those predicted, as is visible from the top panel of
Fig.~\ref{fig:c2_dyn_scaling}. In particular, we note that the
correspondence between theory and simulation is best for small times
(compared to $\tau_\mathrm{sat}$).

A final comment on dynamic scaling and critical behaviour.  Usually
near critical points, the behaviour is determined only by long
lengthscales, the symmetries of the Hamiltonian, and 
dynamically conserved quantities.
In the plaquette model, the short
lengthscales associated with the underlying lattice are relevant even
at very small temperatures, as can been seen by the strongly
anisotropic behaviour of $C_{2,2}(x,y,t)$. However
the vanishing of the static correlation function, $C_4^{\rm stat}
(x,y)$, away from the axes of the underlying lattice is a result of the
symmetries of the Hamiltonian. The similar reduction of
$C_{2,2}(x,y,t)$ has its origin in a conserved quantity, namely the
number of excited plaquettes in every row and column of the square
lattice which is conserved modulo 2.  This is the reason for the
confinement of dimers to one dimension. The absence of any such
conserved quantiy in the froth model explains why it is so different
from the plaquette model.

\subsection{Suppression of diffusion in the plaquette model}
\label{subsec:aging}

This subtle difference between froth and plaquette
models can be made more spectacular when an aging situation
is considered. 
It is known~\cite{Davison} that the aging
behaviour of the froth model after a quench from high temperature is
consistent with diffusion and annihilation of defects, $A+A\to 0$,
and has therefore an energy density that
decays as $(\log t)/t$ in two dimensions.

\begin{figure}
\epsfig{file=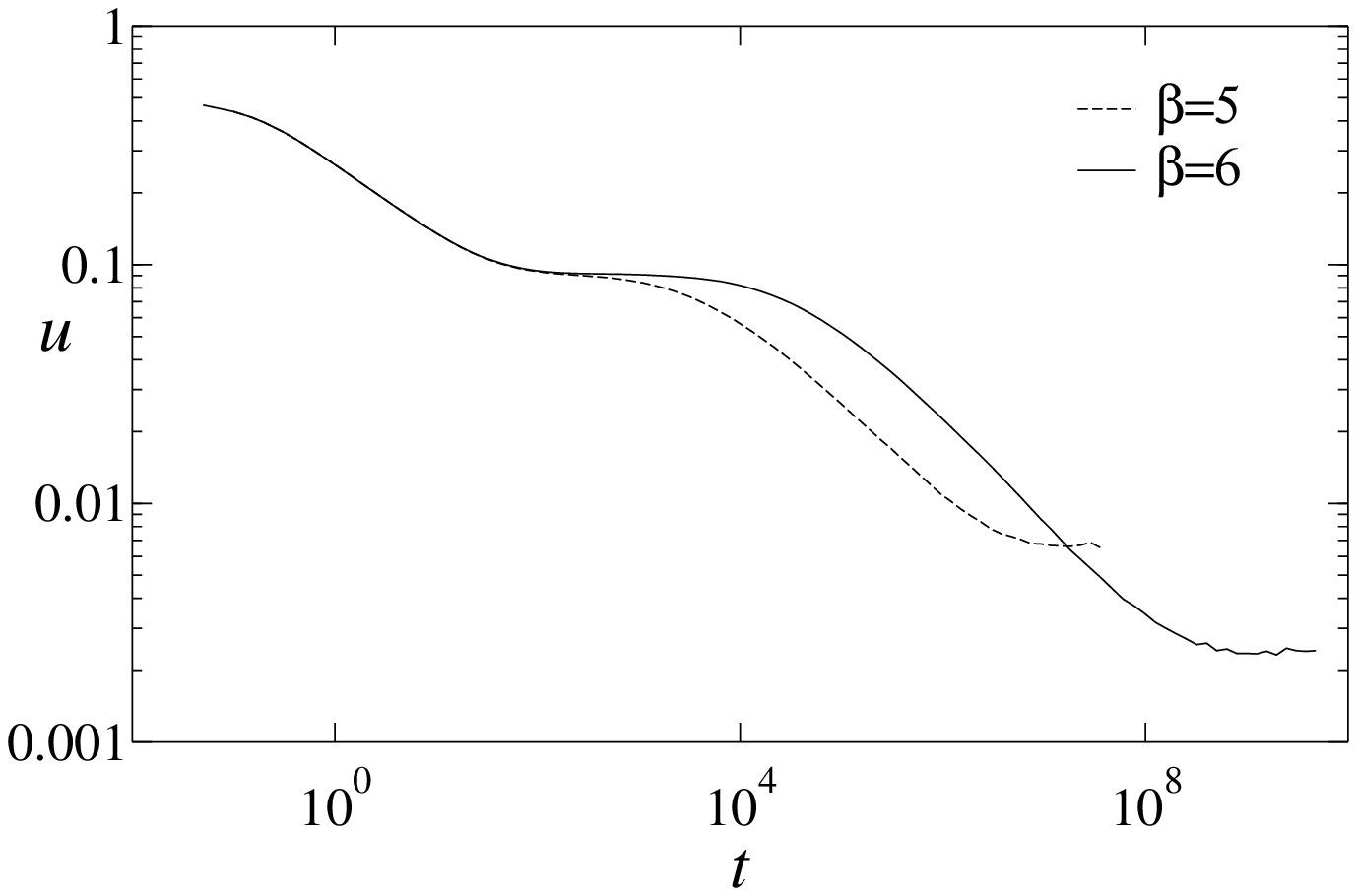,width=0.95\columnwidth}
\epsfig{file=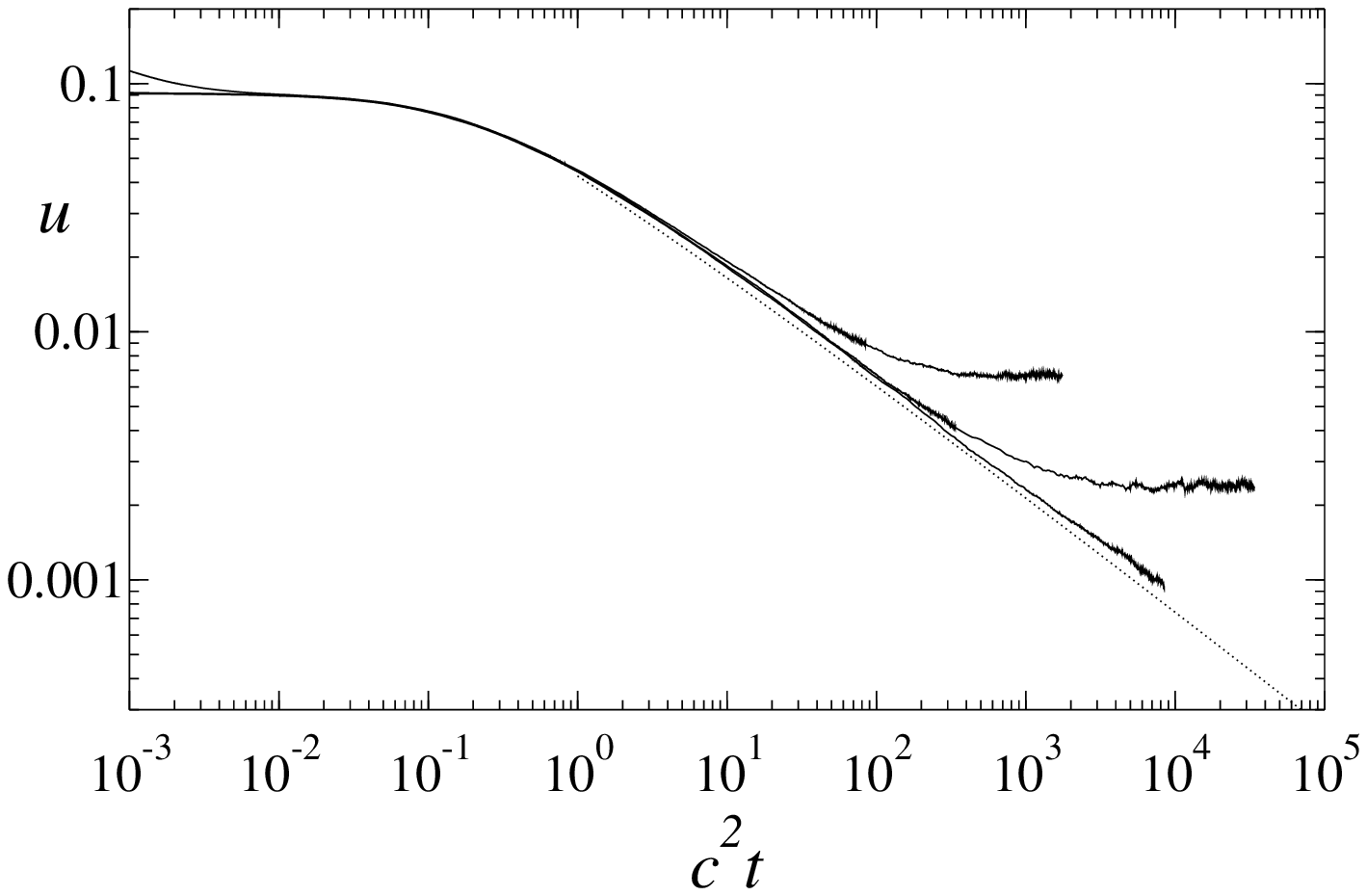,width=0.95\columnwidth}
\caption{Time decay of the density of defects, $u$, 
after a quench from infinite temperature. (Top)
Data for $\beta=5$ and $\beta=6$. 
The initial plateau at short times is a signature of the onset of 
activated dynamics, common to many facilitated spin models. As the 
density of excited plaquettes falls, the system enters a state in which 
the vast majority of spins must overcome an energy barrier in order to 
flip. Thus the dynamics slows down, see~\cite{review}.
The final plateau is equilibration at a density $(1+e^{\beta})^{-1}$.
(Bottom) Quench data from long times, at $\beta=5,6,7$; these collapse
as a function of rescaled time $c^2t$,
until such time as the system equilibrates. The dotted line is 
a fit: $A\sqrt{\log(Bc^2t)/(c^2 t)}$ with $(A,B)=(0.02,90)$.}
\label{fig:quench}
\end{figure}

In the plaquette model, the conservation of the parity of the number
of excitations in each row or column implies that a single defect
cannot simply diffuse.  Accordingly, we show in Fig.~\ref{fig:quench}
that the aging behaviour of the plaquette model is moved out of a
simple annihilation-diffusion regime, and the energy density decays
much more slowly than $(\log t)/ t$. In Ref.~\cite{Jack8v2004cm}, we
showed that this decay can also be fitted satisfactorily by a power
law $t^{-0.45}$ (arising from an decay equation $\partial_t u \propto
u^{2.2}$ where we attributed the anomalous exponent to a fluctuation
correction).  We will argue below that the fluctuation correction is
in fact logarithmic; we also note that the same decay was recently
fitted on a much smaller time window as $t^{-1/3}$~\cite{Espriu2004}, 
invalidated by our simulations performed on much larger timescales. 

To explain this behaviour, we recall that 
in two-dimensional annihilation-diffusion processes
the effective rate equation is
\begin{equation}
u  \sim \frac{\log t}{Dt} + \mathcal{O}(t^{-1}),
\label{equ:rho_decay}
\end{equation}
where $u=\langle H/N + 1/2 \rangle$ is the density of excited
plaquettes, and $D$ is the diffusion constant.  In the plaquette model
any diffusive step in fact involves movement of at least two excited
plaquettes. The simplest such horizontal moves involve isolated
excited plaquettes in either the same row, or in adjacent rows, as in
Fig.~\ref{fig:dimer_sketch}, where one isolated defect has moved by
emitting a dimer.  This pair must then be absorbed by the other
isolated excited plaquette, or the system will revert to its original
configuration. If the separation of the two isolated excitations is $d
\gg 1$ then the probability of the dimer traveling across the gap is
approximately $d^{-1}$~\cite{Espriu2004}.  The diffusion constant is
therefore proportional to $e^{-2\beta} \langle d^{-1} \rangle$ where
the exponential prefactor arises from the activation barrier to dimer
creation.

In equilibrium one has $\langle d^{-1} \rangle \sim 1/\xi_2^{\rm stat}
\sim c$. The diffusion constant and inverse relaxation time are both
proportional to $c^{-3}$. Out of equilibrium we should also consider
correlations in the density of excited plaquettes. However these
correlations take place on a lengthscale $\approx u^{-1/2}$, which is
much smaller than the likely values of $d$. Therefore we neglect these
fluctuations to get $D \propto u$.  Substituting into
Eq.~(\ref{equ:rho_decay}), we find that
\begin{equation}
u \sim \sqrt{\frac{\log t}{t}}.
\end{equation}
This prediction is in reasonable agreement with Fig.~\ref{fig:quench}. However
very many decades of time would be necessary to confirm 
beyond any doubts this particular form 
of logarithmic corrections.

\subsection{Dynamical correlations in the froth model}
\label{subsec:dyn_froth}

Having described the dynamical correlations of the plaquette model in
some detail, we can understand the correlations in the froth model
rather easily. There is no restriction on defect concentration in rows
and columns of the hexagonal lattice, so the anisotropic behaviour of
the plaquette model is absent. However the phenomenology is a simple
generalisation of the behaviour described above.

In the absence of spin degrees of freedom, we consider the
persistence function $b_{ij}(t,t')$ instead of the autocorrelation function
as a local dynamical function.
That is, we define 
\begin{equation}
C'_{2,2}(x,y,t)=\frac{\langle b_{ij}(t,t_0) b_{i+x,j+y}(t,t_0)\rangle -
\langle b_{ij}(t,t_0) \rangle^2}{
\langle b_{ij}(t,t_0)\rangle -\langle b_{ij}(t,t_0) \rangle^2},
\end{equation} 
where the prime indicates that we deal with persistence rather than
autocorrelation functions.
In the plaquette model, the scaling of $C'_{2,2}(x,y,t)$ is very
similar to that of $C_{2,2}(x,y,t)$, compare
Figs.~\ref{fig:pers_space} and~\ref{fig:ac_sketch}.  Therefore it is
sensible to compare $C'_{2,2}$ for the froth model with $C_{2,2}$ for
the plaquette model.  Further, since we will find $C'_{2,2}$ to be
isotropic for all but the smallest lengthscales in the froth model, we
write its circular average as $C'_{2,2}(r,t)$ and use this latter
correlation function in what follows.

To generalise the arguments of section~\ref{subsec:scaling} to 
the froth model, we first recall that for random walkers in 
two dimensions, the typical distance 
traveled in time $t$ scales as $t^{1/2}$, but the typical
number of sites visited by the walker scales as $t/\log t$.
Logarithmic corrections arise because  
$d=2$ is a marginal dimension for free random walks, 
see Ref.~\cite{crist} and Refs. therein.

We expect therefore the following generalisation of the three regimes
observed in the plaquette model. At small times we expect the
lengthscale to grow as $\xi_{2,2}' \sim t^{1/2}$ and the
susceptibility to grow with the number of visited sites, $\chi_{2,2}'
\sim t/ \log t$.  Saturation occurs at a lengthscale controlled by the
mean free path of diffusing dimers, which scales as $c^{-1/2}$ and
represents the separation of free excitations which absorb the
dimers. We note that this lengthscale is smaller than the mean free
path of the plaquette model, which was the typical separation between
single defects in the same row or column of the square lattice.  Since
the scaling is diffusive, saturation will occur at a time,
$\tau_\mathrm{sat}' / \log \tau_\mathrm{sat}' \sim c^{-1}$.  The decay
of the susceptibility will finally be controlled by the decay of the
persistence function on a timescale given by $\tau / \log \tau \sim
c^{-2}$.

\begin{figure}
\epsfig{file=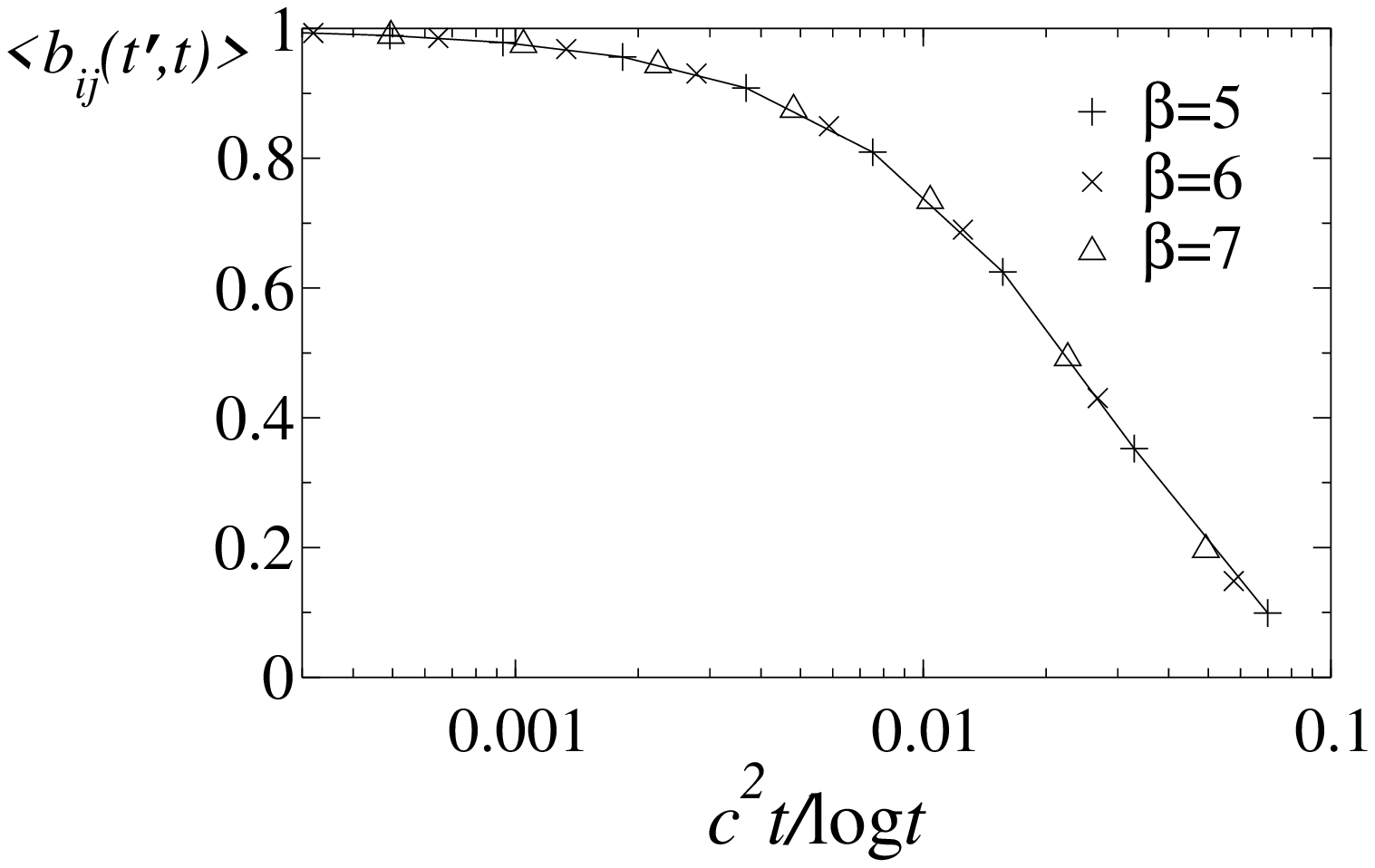,width=0.9\columnwidth}
\epsfig{file=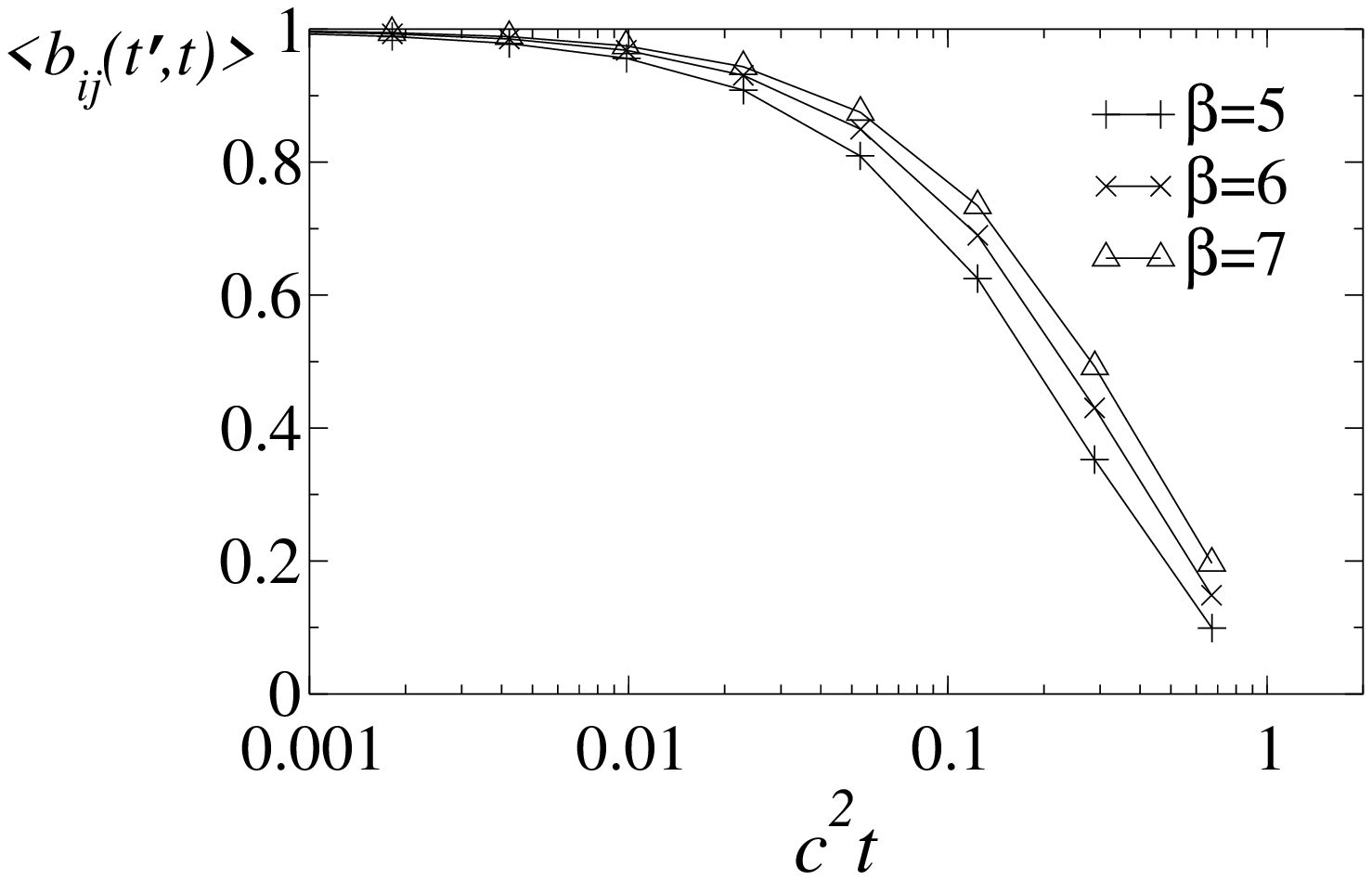,width=0.9\columnwidth}
\caption{Two scaling plots of the persistence function in the froth
model, showing the effect of logarithmic corrections in two
dimensions}
\label{fig:froth_pers_t}
\end{figure}

To illustrate these predictions, we first
show the persistence function in Fig.~\ref{fig:froth_pers_t}, 
plotted as a function of the 
naively rescaled variable $c^2 t$, and also against
the more appropriate variable
$c^2 t/\log t$. Since we work over only a few decades of time,
we may also obtain reasonable fit to a form $c^\delta t$, but the 
logarithmically corrected variable which has no free parameters 
appears indeed as the appropriate one.

\begin{figure}
\epsfig{file=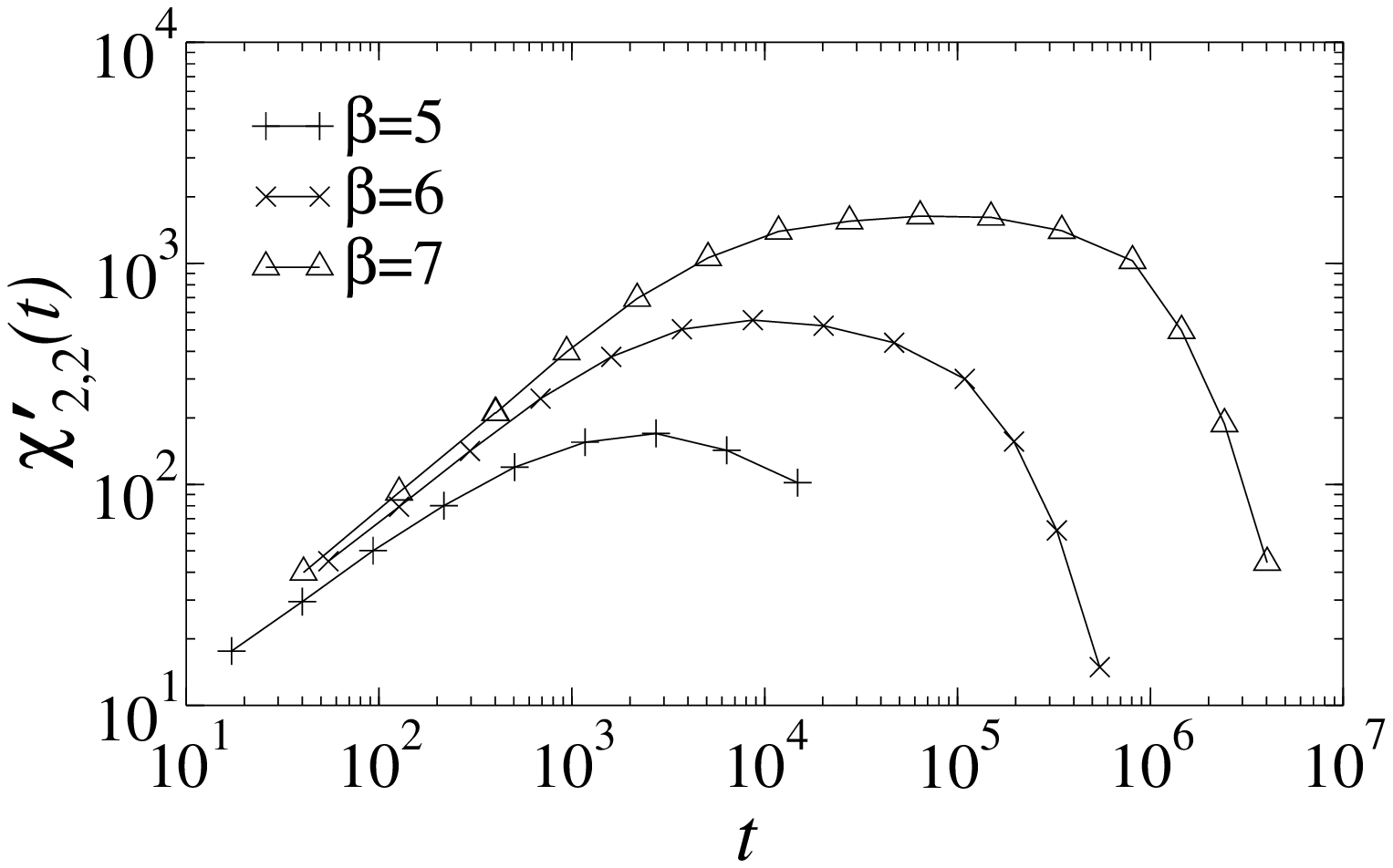,width=0.85\columnwidth}
\epsfig{file=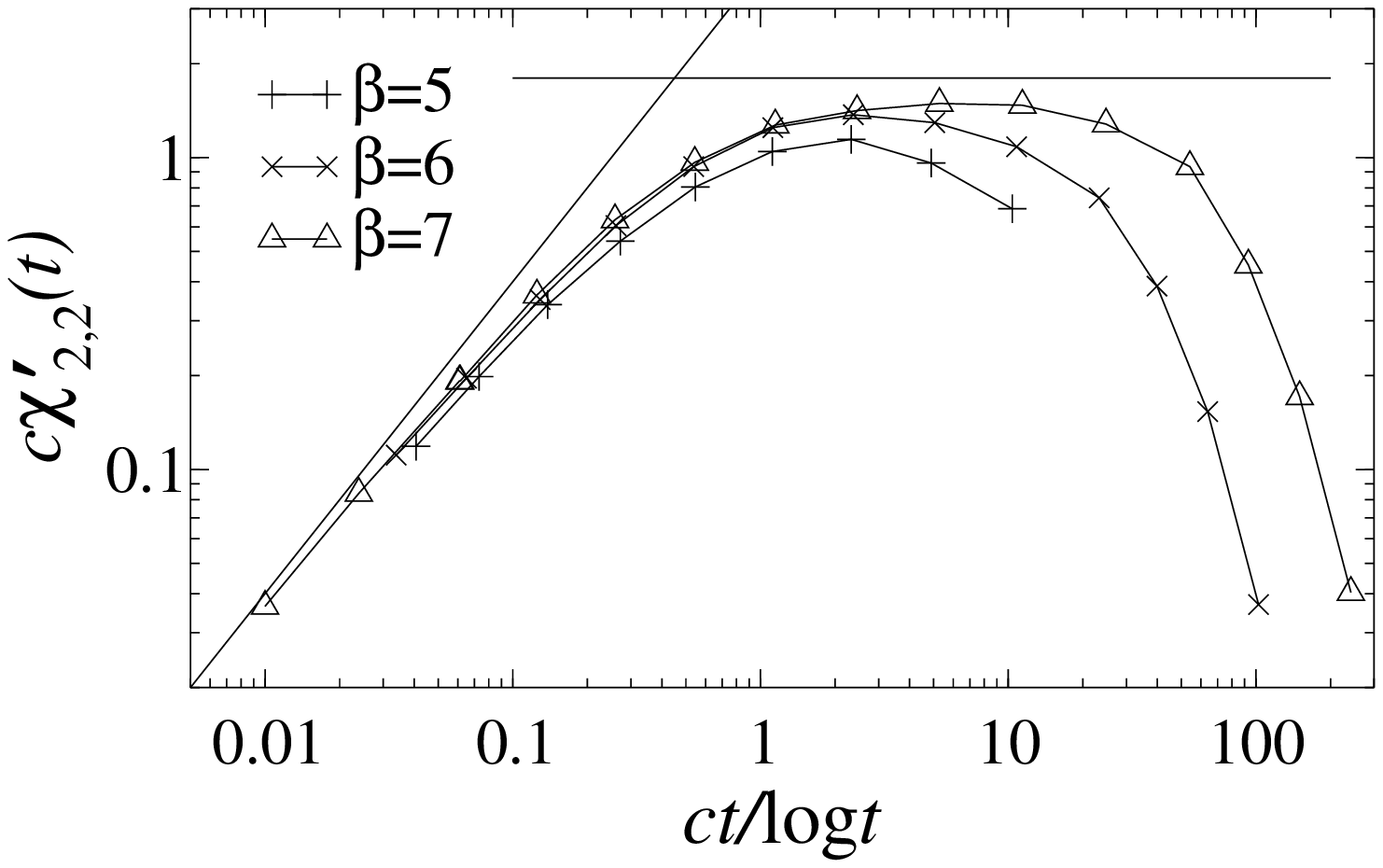,width=0.85\columnwidth}
\epsfig{file=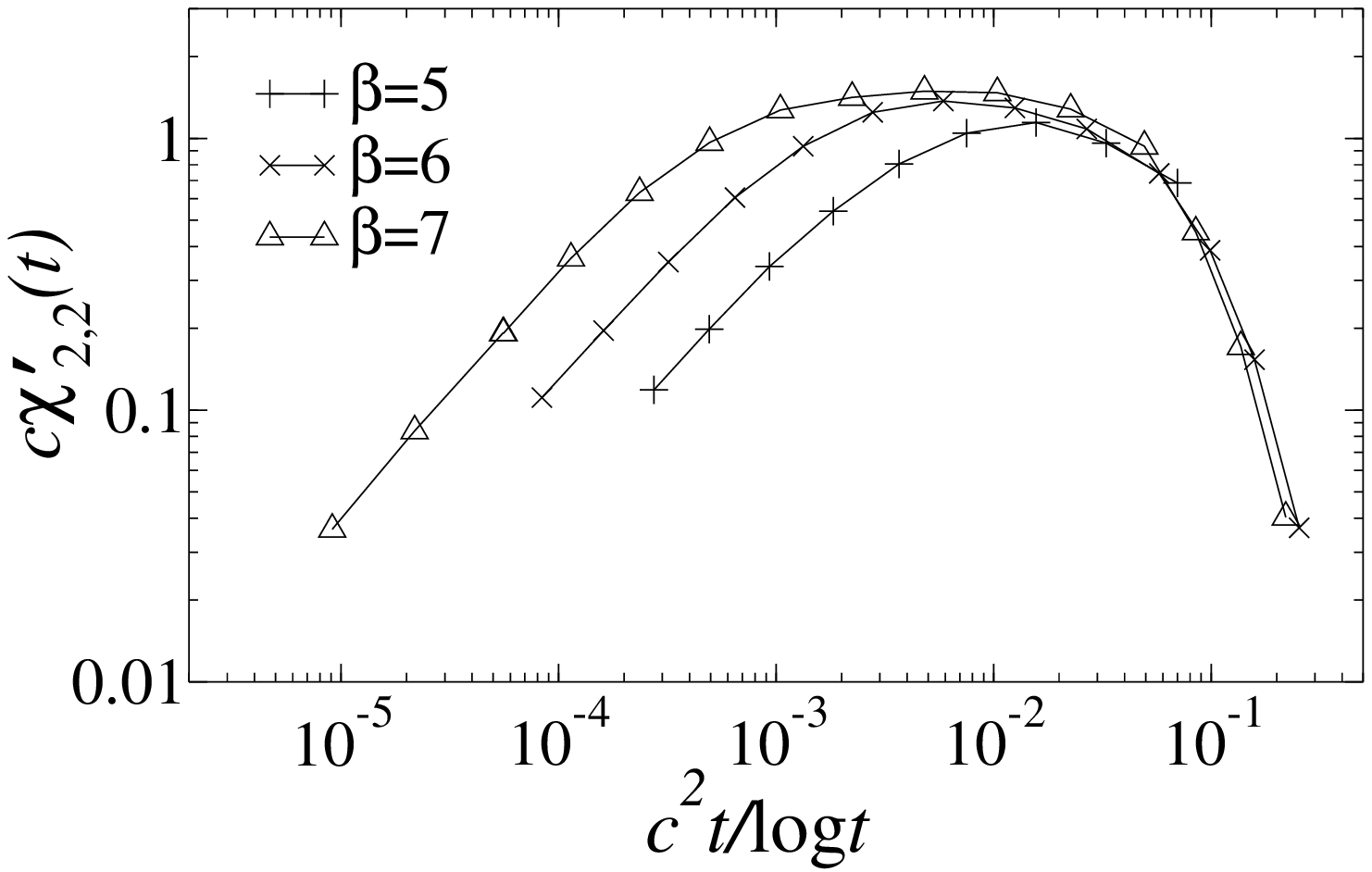,width=0.85\columnwidth}
\caption{
Top: Plots of $\chi_{2,2}'(t)$ at various temperatures. 
Middle: Rescaling of time to $c t/ \log t$ showing
saturation at $c \tau_\mathrm{sat} / \log \tau_\mathrm{sat}\sim 1$. 
Lines are the random walk prediction
$\sim t/\log t$ and a horizontal plateau showing 
saturation at $c \chi_{2,2}' \sim 1.8$. 
Bottom: Rescaling of time to $c^2 t / \log t$ showing decrease of correlations
which scales as the persistence function in Fig.~\ref{fig:froth_pers_t}.
} \label{fig:froth_chi}
\end{figure}

In Fig.~\ref{fig:froth_chi} we show the susceptibility 
associated with the persistence function
\begin{equation}
\chi_{2,2}'(t) = \sum_{xy} C'_{2,2}(x,y,t).
\end{equation}
Saturation takes place at a value of $\chi_{2,2}'$ that is
proportional to $c^{-1}$ and the approach to saturation is a universal
function of $ct/\log t$.  The susceptibility then decays on the
timescale of the persistence function.

The collapse of the persistence and of the susceptibility 
with a single rescaled time variable 
is consistent with their dependence on the number
of sites visited by each random walker. When considering spatial
correlations, the situation is slightly more complicated. While the
relevant areas scale as $t/\log t$, the distances traveled by
walkers are distributed as universal functions of $r^2/t$.
We therefore expect that while the susceptibility varies as 
a scaling function of $t/\log t$, the moments of the probability
distribution should have a normal diffusive behaviour. That is, we expect
\begin{equation}
\chi_{2,2}' (t) = \int 2\pi r \mathrm{d}r \, C'_{2,2}(r,t) \sim t/\log t, 
\end{equation}
but simultaneously
\begin{equation}
\frac{\int 2\pi r \mathrm{d}r \, r^n C'_{2,2}(r,t)}{\int 2\pi r \mathrm{d}r \,
C'_{2,2}(r,t)} \sim t^{n/2}
\label{equ:froth_c22_scaling}
\end{equation}
These relations were proved recently in Ref.~\cite{crist}
for non-interacting random walkers in two dimensions 
and unnormalized dynamic spatial correlators.
An alternative statement of Eq.~(\ref{equ:froth_c22_scaling})
is that 
\begin{equation}
C'_{2,2}(r,t) = \frac{\chi_{2,2}'(t)}{t} \, f_3\left( \frac{r^2}{t}
\right), \qquad t\ll \tau_\mathrm{sat},
\label{equ:froth_c22_dyn_coll}
\end{equation}
which should be valid for $r\gg1$. This scaling relation must break
down at small distance since both $C_{2,2}'(0,t)$ and $\chi_{2,2}'(t)$
are scaling functions of $t/ \log t$, incompatible with
(\ref{equ:froth_c22_dyn_coll}).

\begin{figure}
\epsfig{file=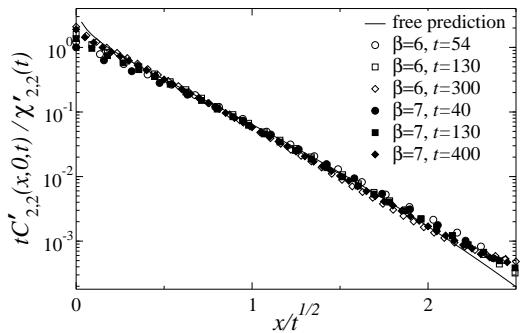,width=0.85\columnwidth}
\caption{Scaling plot of $(t C_{2,2}'/\chi_{2,2}')$ as 
a function of the reduced variable $r^2/t$.
The full line is the independent random walk prediction
(\ref{eq:crist}) derived in Ref.~\cite{crist}. The fit
is valid for $r\gg1$ so the deviations at small $r$ are
greatest at small $t$. The deviations at large $r$ are of the order
of the numerical uncertainty in our simulations, but might also
be an effect of the zig-zag motion of the dimers.}
\label{fig:froth_cr_unsat}
\end{figure}

We test the collapse of Eq.~(\ref{equ:froth_c22_dyn_coll}) in
Fig.~\ref{fig:froth_cr_unsat}. There are tiny deviations at small
$r^2/t$ as expected. Moreover we get reasonable agreement with the
analytical results of Ref.~\cite{crist} which predict
\begin{equation}
\frac{t C_{2,2}'(r,t)}{\chi_{2,2}'(t)} = 
\frac{1}{2\pi D}
 \int_1^\infty \mathrm{d}x \,
\left( \frac{1}{x} - \frac{1}{x^2} \right)
e^{-xr^2/2Dt} ,
\label{eq:crist}
\end{equation}
where $D$ is the diffusion constant of the random walkers. The
diffusion constant for dimers in the froth model will be of order
unity since the moves carry no energy penalty. In
Fig.~\ref{fig:froth_cr_unsat} we show reasonable fit with $D=1$.  We
note that the dimers diffuse in a zig-zag fashion. A given dimer makes
random steps of length unity along two non-orthogonal directions that
depend on the relative position of the two defects that form the dimer
(this orientation is constant as the dimer diffuses).  This zig-zag
motion will lead to a reduction in the diffusion constant.  Since
there are only three possible dimer orientations, there is also the
possibility of anisotropy in the correlations, but this was not
observed in our simulations. Since the scaling of
Fig.~\ref{fig:froth_cr_unsat} is good and the fit reasonable, we do
not attempt any more rigorous analysis of these effects.

\begin{figure}
\epsfig{file=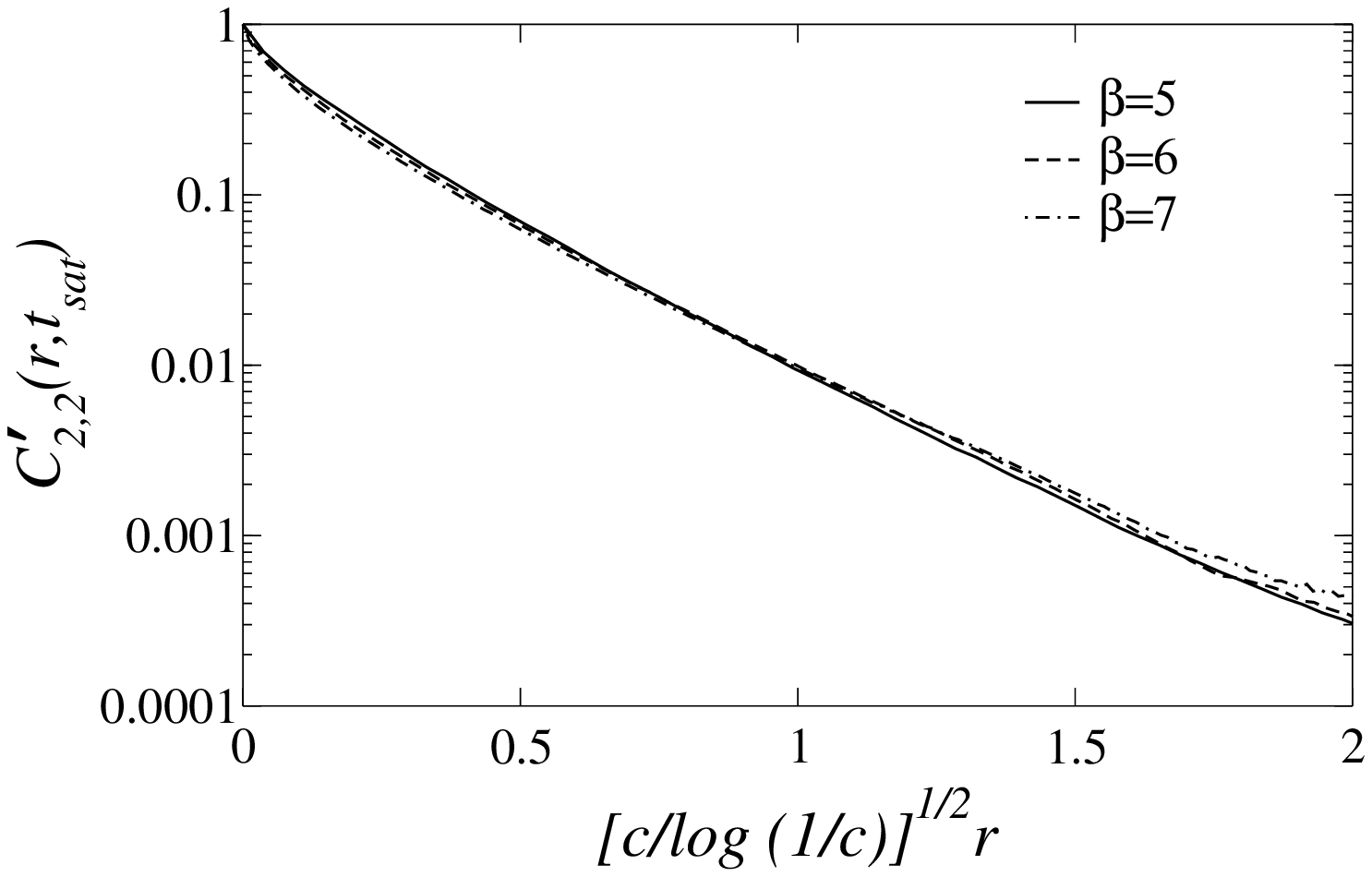,width=0.85\columnwidth}
\epsfig{file=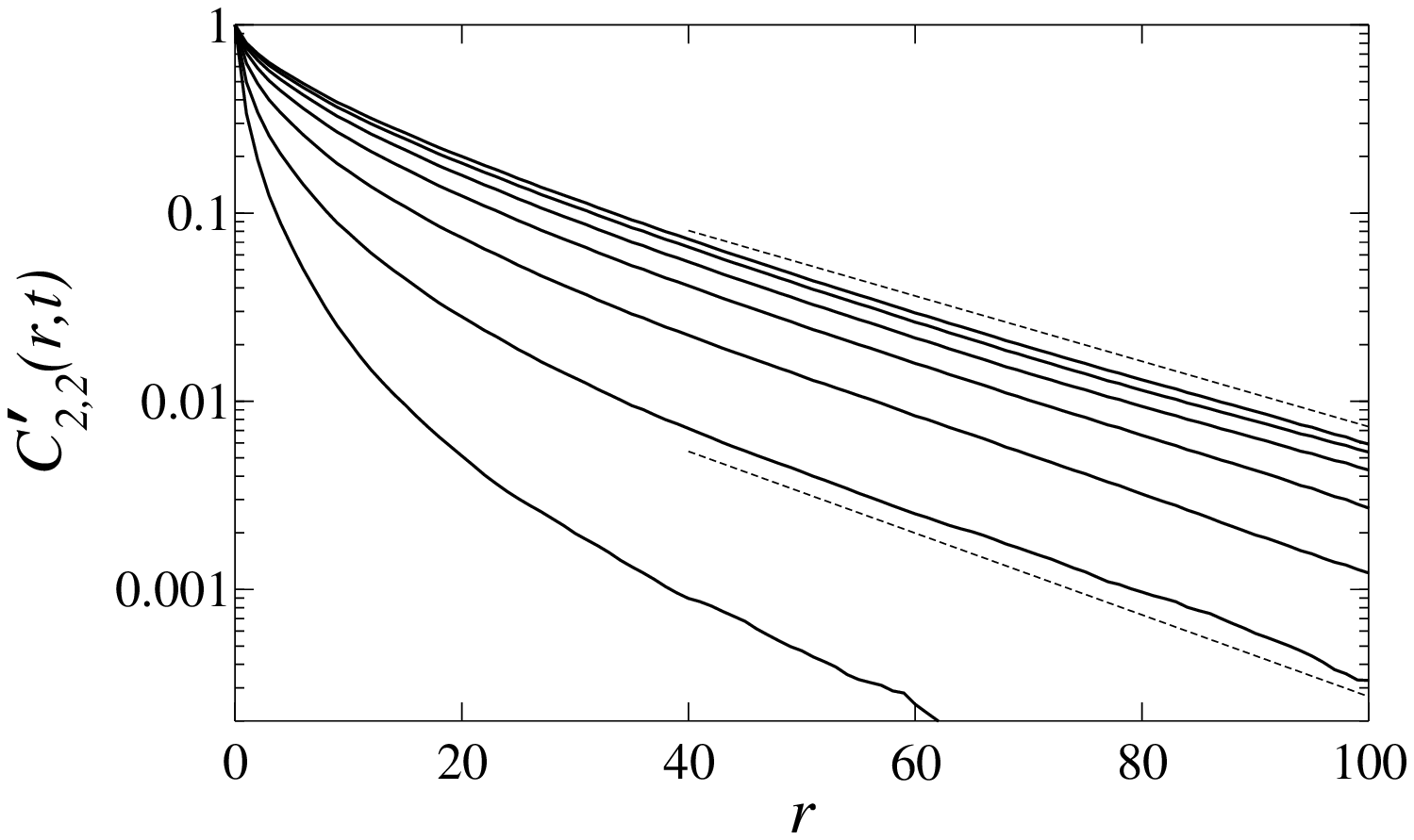,width=0.85\columnwidth}
\caption{Top: Plot of $C_{2,2}'$ in the state with a saturated
correlation area showing good scaling as a function of
$r\sqrt{c\log(1/c)}$.  Bottom: Plot of $C_{2,2}'(r)$ 
at logarithmically spaced times 
between $0.3\tau$ and $10\tau$, with $\beta=7$.
As time increases, the
strength of the correlations decreases by two orders of magnitude. 
On the other hand, the gradient of these traces at 
large distances gives the correlation length
which is decreasing rather slowly with time (we show dashed 
lines marking $e^{-r/25}$ and $e^{-r/20}$). We constrast this behaviour
with the unsaturated regime in which the correlations are always strong,
and the correlation length \emph{increases} as $t^{1/2}$. }
\label{fig:froth_crsat}
\end{figure}

Finally we consider the behaviour of the dynamic lengthscale at
saturation.  The number of sites visited by a walker increases as
$(t/\log t)$. Since the absorbing sites are uniformly distributed the
walker will be absorbed on a timescale satisfying $\tau_\mathrm{sat}
\sim c^{-1} \log \tau_\mathrm{sat}$.  In that time the walker travels
an average distance
\begin{equation}
\xi_\mathrm{sat} \sim \tau_\mathrm{sat}^{1/2} \sim 
c^{-1/2} [ \log c^{-1} + \mathcal{O}(\log \log c^{-1}) ]^{1/2}
\end{equation}
In Fig.~\ref{fig:froth_crsat} (top) we present scaling of the
saturated correlation function according to this law.  We also show in
Fig.~\ref{fig:froth_crsat} (bottom) that when times become of the
order of or larger than the persistence time, the change in the
susceptibility $\chi_{2,2}'(t)$ is mainly due to a change in the
strength of the correlations while the dynamic lengthscale
$\xi_{2,2}'$ is approximately constant, just as in the plaquette model.  
We attribute the weak decrease of the correlation length at the longest 
times to the fact that the persistence function retains rather little
information for times much greater than the relaxation times $\tau$.
As a
consequence the free random walk calculations in Ref.~\cite{crist} do
not apply at large times because they do not take saturation effects
into account.

To summarise the results of this subsection, we have shown that the
froth model behaves similarly to the plaquette model with a mean free
path for dimers that scales as $c^{-1/2}$ and a relaxation timescale
that scales as $c^{-2}$ with important logarithmic corrections due to
the dimensionality of the random walks performed by freely diffusing
dimers.

\section{Conclusion}
\label{sec:conc}

The plaquette model is a spin model with simple dynamics and no finite
temperature thermodynamic singularities. Its Hamiltonian gives rise to
effective kinetic constraints, and therefore to dynamical
heterogeneity.  Despite simple thermodynamics in the plaquette
representation, spins have static correlations whose lengthscales
diverge at low temperature. The symmetries of the Hamiltonian mean
that only rather specific correlation functions have non-zero
expectations. However, measurement of the fluctuations in two-point
quantities allow extraction of the relevant lengthscales.

Spatial correlations are strongly anisotropic. For static
correlations, this is the result of a symmetry of the spin system.
Dynamical correlations are anisotropic because of a microscopic
conservation law.  We have identified three lengthscales in the
plaquette model. The mean spacing between excited plaquettes,
$\xi_4^{\rm stat} \sim c^{-1/2}$ controls four point static
correlations. Fluctuations in two-point structure factors are
controlled by $\xi_2^{\rm stat} \sim c^{-1}$ which represents the mean
distance between adjacent excited plaquettes in the same row or
column.  Finally there is a dynamical lengthscale $\xi_{2,2}(t)$ whose
behaviour is diffusive at small times and saturates to $\xi_2^{\rm
stat}$ at larger times.

We also identified two timescales in the system. The first is the
saturation time, $\tau_\mathrm{sat}\sim c^{-2}$, which separates the
regime in which dynamical correlations have an increasing lengthscale
from the regime in which their lengthscale is saturated.  The second
is the relaxation time $\tau\sim c^{-3}$ which represents the typical
time for the spin field to lose the memory of its initial
configuration. This separation of timescales arises from the
non-trivial conserved quantities at the zero temperature dynamical
fixed point. We also showed that these effects lead to unusual aging
behaviour.

We explained the dynamical length and timescales in the plaquette
model in terms of diffusion of excitation pairs along one-dimensional
paths. We showed that the froth model correlations are obtained by
generalising these results to allow the excitation pairs to diffuse in
two dimensions. This introduces logarithmic correlations, but once
these have been taken into account the scaling is as expected for
non-interacting random walkers.

The results connecting static and dynamic correlations in the
plaquette model demonstrate one specific mechanism by which a
`mobility field' emerges from a more familiar interacting spin
field. They are relevant to the description of structural glasses in
these terms. The fact that all of the dynamical correlations can be
explained in terms of independent random walkers provides further
evidence that this behaviour is rather universal in systems with
dilute diffusing defects.

\acknowledgments
 
We thank G.~Biroli and C.~Toninelli for discussions. 
This work was supported by CNRS (France), EPSRC Grants No.\
GR/R83712/01 and GR/S54074/01, and University of Nottingham Grant No.\
FEF 3024.

\end{document}